\documentclass[useAMS,usenatbib]{mn2e}
\usepackage{epsf,rotating}
\usepackage{amsmath}
\usepackage{subfigure}
\usepackage{lscape}

\usepackage{natbib}

\newcommand{\nh}{n_{\rm H}}
\newcommand{\lya}{Ly$\alpha$}

\newcommand{\hmpc}{h^{-1}{\rm Mpc}}

\newcommand{\kms}{\;{\rm km}\,{\rm s}^{-1}}
\newcommand\cdunits{{\rm cm}^{-2}}

\newcommand{\gad}{{\sc Gadget-2}}
\newcommand{\autovp}{{\sc AutoVP}}
\newcommand{\ion}[2]{\hbox{#1\,{\sc #2}}}
\newcommand{\vw}{v_{\rm w}}

\newcommand{\apj}{ApJ}
\newcommand{\mnras}{MNRAS}
\newcommand{\apjs}{ApJS}

\newcommand{\msolar}{\;{\rm M}_{\odot} }

\newcommand{\mlo}{\;{${10}^{11}M_\odot$}}
\newcommand{\mmid}{\;{${10}^{12}M_\odot$}}
\newcommand{\mhi}{\;{${10}^{13}M_\odot$}}

\title[Circum-Galactic Absorption in Simulations]{Hydrogen \& Metal Line Absorption Around Low-Redshift Galaxies in Cosmological Hydrodynamic Simulations}

\author[Ford et al.]{
\parbox[t]{\textwidth}{\vspace{-1cm}
Amanda Brady Ford$^1$, Benjamin D. Oppenheimer$^2$, Romeel Dav\'{e}$^1$, Neal Katz$^3$, Juna A. Kollmeier$^4$, David H. Weinberg$^5$}
\\\\$^1$ Astronomy Department, University of Arizona, Tucson, AZ 85721, USA
\\$^2$ Leiden Observatory, Leiden University, PO Box 9513, 2300 RA Leiden, Netherlands
\\$^3$ Astronomy Department, University of Massachusetts, Amherst, MA 01003, USA
\\$^4$ Observatories of the Carnegie Institution of Washington, Pasadena, CA 91101, USA
\\$^5$ Astronomy Department and CCAPP, Ohio State University, Columbus, OH 43210, USA
}

\begin{document}

\pubyear{2012}

\maketitle

\label{firstpage}

 \begin{abstract}
We study the physical conditions of the circum-galactic medium (CGM)
around $z=0.25$ galaxies as traced by \ion{H}{i} and metal line
absorption, using cosmological hydrodynamic simulations that include
galactic outflows.  Using lines of sight targeted at impact parameters
from 10~kpc to 1~Mpc around galaxies with halo masses from
$10^{11}-10^{13}M_\odot$, we study the physical conditions and their
variation with impact parameter $b$ and line-of-sight velocity
$\Delta v$ in the CGM as traced by \ion{H}{i}, \ion{Mg}{ii},
\ion{Si}{iv}, \ion{C}{iv}, \ion{O}{vi}, and \ion{Ne}{viii} absorbers.
All ions show a strong excess of absorption near galaxies compared
to random lines of sight. The excess continues beyond 1~Mpc,
reflecting the correlation of metal absorption with large-scale
structure.  Absorption is particularly enhanced within about $\Delta
v<300 \kms$ and roughly 300~kpc of galaxies (with distances somewhat
larger for the highest ion), approximately delineating the CGM;
this range contains the majority of global metal absorption.  Low
ions like \ion{Mg}{ii} and \ion{Si}{iv} predominantly arise in
denser gas closer to galaxies and drop more rapidly with $b$, while
high ions \ion{O}{vi} and \ion{Ne}{viii} trace more diffusely
distributed gas with a comparatively flat radial profile; \ion{C}{iv}
is intermediate.  All ions predominantly trace $T\sim 10^{4-4.5}K$
photo-ionised gas at all $b$, but when hot CGM gas is present (mostly
in larger halos), we see strong collisionally-ionised \ion{O}{vi}
and \ion{Ne}{viii} at $b\leq 100$ kpc.  Larger halo masses generally
produce more absorption, though overall the trends are not as strong
as that with impact parameter.  These findings arise using our favoured
outflow scalings as expected for momentum-driven winds; with no
winds, the CGM gas remains mostly unenriched, while our outflow
model with a constant velocity and mass loading factor produce
hotter, more widely dispersed metals.
\end{abstract}

\section{Introduction}

The intergalactic medium (IGM) contains most of the cosmic baryons
at all epochs (e.g., \citealp{dav01}).  It also contains a substantial
fraction of cosmic metals~\citep[e.g.,][]{fer05,dav07,wie09b},
products of star formation within galaxies that have been dispersed
out of galaxies presumably via large-scale galactic
outflows~\citep[e.g.,][]{agu01b,opp06}.  The distribution and
physical state of these metals thus provide a powerful tracer of
galactic outflow processes, which are a key uncertainty in current
galaxy formation models.

\citet{opp12} argued from cosmological simulations with outflows
that the IGM is enriched in an ``outside-in" manner, with low-density
regions enriched at earlier epochs, and that subsequently produced
metals tend to remain closer to galaxies over the last 10~Gyr of
cosmic time.  In simulations by \citet{opp08} that broadly reproduce
IGM enrichment observations, outflowing metals reach a median
distance of roughly 100 physical kpc before turning around, which
is well outside the virial radius of small early galaxies but results
in ``halo fountains" within larger, present-day systems.  Hence by
low redshifts, a large fraction of the ejected metals are expected
to reside close to galaxies, within what is now frequently called
the circum-galactic medium (CGM).  Indeed, observations indicate
that \ion{C}{iv}, \ion{Mg}{ii} and \ion{O}{vi} absorbers can often
be associated with galaxies within tens or hundreds of
kpc~\citep[e.g.,][]{ste92,sto06,wak09,pro11,che01,che09}. The CGM
thus provides a fossil record of early IGM enrichment combined with
the gravitational growth of structure, in addition to metals deposited
by more recent outflows.

Recent observations using {\it Hubble's} Cosmic Origins Spectrograph
(COS) provide a greatly enhanced view of CGM metals at low redshifts.
\citet{tum11} showed that the CGM around star-forming galaxies
contains an amount of oxygen comparable to or possibly exceeding that
in the interstellar medium (ISM) of those galaxies, thereby directly
demonstrating that metals have been expelled {\it en masse} from the
ISM~\citep{tri11}.  The sensitivity of COS not only permits the selection
of more numerous, fainter
background sources that can probe the CGM of low-redshift galaxies,
but also enables the detection of weaker and less common ions such
as \ion{Ne}{viii} and \ion{Si}{iv}.  These data will provide new and
stringent constraints on theoretical models of outflows that connect
galaxies and their surrounding gas.

Cosmological hydrodynamic simulations offer a uniquely well-suited
platform to study the co-evolution of galaxies and their surrounding gas.
Such simulations dynamically incorporate galactic outflows driven by
star formation, thereby self-consistently enriching the IGM and CGM with
the by-products of stellar death, as well as accounting for the movement and
physical state of such metals driven by hierarchical structure formation.
With simulations of representative cosmological volumes, one can test outflow models by comparing the
predicted statistics directly to absorption line observations of
\ion{H}{i} and metals, around galaxies and in the diffuse IGM. Tracking the complex inflow and outflow processes that
govern both galaxies and the IGM is necessary to fully understand how
IGM and CGM enrichment occur, and to properly interpret present and
upcoming data with COS and other facilities.

In this paper we continue our exploration of \ion{H}{i} and metal
absorption in the low-redshift IGM using cosmological hydrodynamic simulations, building on the work presented by  \citet{dav10} and
\citet{opp12}.   In those
papers we studied absorption properties along random lines of sight
through cosmological volumes, which only occasionally intercepted the
CGM of galaxies.  In this work, we select targeted lines of sight at
various impact parameters around galaxies in our simulations to directly
probe the absorption properties and physical conditions of gas within
the CGM.  The goal of this paper is to understand the relationship
between galaxy properties and HI and metal absorbers that probe the gas
around those galaxies. 

We explore a range of metal ionisation potential states from
low (\ion{Mg}{ii}) to high (\ion{Ne}{viii}), as well as \ion{H}{i},
and find that the behaviour of many CGM absorption line properties
shows distinct trends with the ionisation level of the tracer species.
CGM absorption lines also show trends with halo mass. These predictions
set the stage for a more detailed understanding of CGM observations and provide a more complete picture for how metals trace the motion of gas in
and out of galaxies. 

Our paper is organised as follows: In \S\ref{sec:sa} we explain
our methods, in \S \ref{sec:pca} we show the density and temperature
of absorbers, in \S\ref{sec:redshift} we explore absorption around
galaxies in redshift space, in \S \ref{sec:b} we study absorption as a
function of impact parameter, in \S \ref{sec:CDD} we examine the column
density distributions as a function of impact parameter and halo mass,
in \S \ref{sec:model} we examine trends with outflow model, and in \S
\ref{sec:conclusions} we present our conclusions.

\section{Simulations \& Analysis}
\label{sec:sa}

\subsection{The Code and Input Physics} 

We use our modified version \citep{opp08} of the N-body+smooth particle
hydrodynamic (SPH) code \gad~\citep{spr05}, which is more fully described
in \S2.1 of \citet{dav10}.  The only code update since \citet{opp08}
is the option to include metal-line cooling rates from \citet{wie09a},
as we discuss below.

Briefly, \gad~computes gravitational forces on a set of particles
using a tree-particle-mesh algorithm and uses an entropy-conserving
formulation of SPH \citep{spr02} to simulate pressure forces and
hydrodynamic shocks.  We include radiative cooling assuming ionisation
equilibrium for primordial species following \citet{kat96} and
metals based on the tables of \citet{wie09a} that assume ionisation
equilibrium in the presence of the \citet{haa01} background.  In
our older simulations, and our ancillary wind model simulations in this paper,
we employ metal-line cooling rates based on the collisional ionisation
equilibrium (CIE) models of \citet{sut93}.  The latter rates
incorrectly over estimate metal-line cooling as photo-ionisation
equilibrium (PIE) suppresses cooling of metal-enriched gas as
demonstrated by \citet{smi11} and \citet{tep11}.  We show
comparisons of these two cooling models in \S\ref{sec:model}.

Star formation follows a \citet{sch59} Law calibrated to the
\citet{ken98a} relation, following \citet{spr03a}.  The interstellar
medium (ISM) is modelled using the analytic sub grid recipe of \citet{mck77}, where supernova energy is returned to ISM particles
using the two-phase SPH formulation of \citet{spr03a}.  Star particles
are spawned from ISM particles probabilistically according to the
instantaneously calculated star formation rate (SFR); an ISM particle
can spawn up to two star particles.

Star formation-driven kinetic feedback is implemented in these
simulations by giving a velocity kick ($\vw$) to ISM particles chosen
probabilistically at a rate proportional to their star formation rate.
The ratio of the mass outflow rate to the star formation rate is termed the
mass loading factor ($\eta$).  We use the relations for $\vw$ and
$\eta$ of momentum-driven winds formulated by \citet{mur05}:
\begin{eqnarray}
\vw &=& 3\sigma\sqrt{{\rm f}_{L}-1}\\ 
\eta&=&\sigma_{o}/\sigma,
\end{eqnarray}
where ${\rm f}_{L}$ is the luminosity in units of the Eddington
luminosity required to expel gas from a galaxy potential, $\sigma_{o}
= 150 \kms$, and $\sigma$ is the galaxy's internal velocity dispersion
\citep{opp08}.  We refer to this wind model as ``vzw" and refer the reader to \citet{opp08} for a more complete description.  We also evolve a model
without winds (``nw") and a constant wind (``cw") model with
$\vw=680\kms$ and $\eta=2$ for all galaxies, as described in \citet{dav10}.

\subsection{Simulation Runs}

Our main model in this paper is a $48 \hmpc$, $2\times 384^3$ simulation
with vzw winds.  This simulation is identical to the r48n384vzw simulation
used by \citet{opp10, opp12} and \cite{dav10,dav11a, dav11b}  except that it uses
\citet{wie09a} metal-line cooling rates.

We choose the vzw wind model as our fiducial model owing to its successes
fitting important properties of galaxies and the IGM at a variety
of redshifts.  At high redshift, simulations with vzw winds provide
adequate fits to the $z=2$ mass-metallicity relationship \citep{fin08,
dav11a}, the $z\ge 6$ galaxy luminosity function \citep{dav07, fin11}, and
observations of IGM metal enrichment at redshifts $\ge 1.5$ \citep{opp06,
opp09b}.  Simulations evolved to $z=0$ reproduce the observed galactic stellar
mass function \citep{bel03,bal08} below ${5}\times{10}^{10}$ ${M}_{\odot}$
\citep{opp10}, various statistical properties of present-day galaxies
\citep{dav11a, dav11b}, and \ion{H}{i} and metal-line statistics from
quasar absorption line (QAL) spectra \citep{opp09, opp12}.  \citet{opp12}
explored a small, $16 \hmpc$, $2\times 128^3$ simulation incorporating
vzw winds and \citet{wie09a} metal-line cooling, finding it to be
their preferred model owing to its theoretically motivated feedback
prescription capable of reproducing the above-listed observations,
combined with a more correct treatment of metal-line cooling. \citet{opp12} also compared the vzw model, as well as a constant wind and no wind model, to observations of random lines of sight, as shown in Figures 6-8, 10, 14-15, 18 and 21. We note
that these simulations do not include any additional mechanisms such
as black hole feedback to quench star formation in massive galaxies
to reproduce the observed galaxy red sequence~\citep[e.g.,][]{gab11}.

\begin{table}
\caption{Comparison of r48n384vzw models.}
\begin{tabular}{lcc}
\hline
Z-Cooling &
CIE$^a$ &
PIE$^b$
\\
\hline
\multicolumn{3}{c}{}\\
$f_{\rm bar}(T\geq 10^5 {\rm K})^c$ &  0.394 & 0.408 \\
$f_{\rm metals}(T\geq 10^5 {\rm K})^d$ &  0.056 & 0.066 \\
$\Omega_{*}/\Omega_{b}^e$ & 0.084 & 0.076 \\
$\rho_{SFR}^f$ & 0.074 & 0.066 \\
dN/dz$(N_{\ion{H}{i}}\geq 10^{14} {\rm cm}^{-2})^g$ & 8.88 & 7.94 \\ 
$\Omega_{\rm OVI}^h$ & 57.6 & 38.8 \\
$\Omega_{\rm CIV}^h$ & 33.9 & 23.7 \\
$\Omega_{\rm SiIV}^h$ & 10.2 & 7.9 \\
\hline
\end{tabular}
\\
\parbox{15cm}{
$^a$Collisional ionisation equilibrium metal cooling rates from \\ \citet{sut93}.  \\
$^b$Photo-ionisation equilibrium metal cooling rates using \\ \citet{wie09a}.\\
$^c$Fraction of diffuse, non-ISM baryons with $T\ge10^5$ K at $z=0.25$.\\
$^d$Fraction of diffuse, non-ISM metals with $T\ge10^5$ K at $z=0.25$.\\
$^e$Fraction of baryons in stars at $z=0.25$.\\
$^f$Instantaneous SFR at $z=0.25$ in $M_{\odot}$ yr$^{-1}$ Mpc$^{-3}$\\
$^g$Frequency of \ion{H}{i} components with $N_{\ion{H}{i}}\geq 10^{14} {\rm cm}^{-2}$ at $z=0.25$.\\
$^h$$\Omega$ summed from all gas outside galaxies at $z=0.25$, in units \\ of $10^{-8}$.\\
}
\label{table:sims}
\end{table}

Because the old r48n384vzw evolved with CIE cooling was featured prominently
in a number of our previous publications \citep{opp10, dav10, dav11a,
dav11b, opp12}, we list the differences in some relevant quantities
relative to using PIE cooling rates in Table \ref{table:sims} at $z=0.25$.
It shows that the differences that result from replacing CIE with PIE
metal-line cooling rates are minor to moderate.  The fraction of baryons
in diffuse (i.e. non-star forming) gas with $T\geq 10^5$ K increases by
3.6\%, while the associated metals increase by 18\%.  Less cooling allows
more baryons and especially metals to remain in the warm-hot intergalactic medium (WHIM), defined as where $T>{10}^{5}$ K and $ \delta < \delta_{th}$, where $\delta_{th}$ is the division between bound and unbound gas \citep{opp12,dav10}. The total integrated star formation (i.e. baryon fraction in stars) is reduced by 10\%, and the instantaneous star formation rate is reduced by 11\%,
both because enriched winds are less capable of cooling and hence
recycling onto galaxies \citep{opp10}.  Considering \lya~absorption
tracing the IGM, the frequency of single Voigt-profile-fitted components
with $N_{\ion{H}{i}}\geq 10^{14} {\rm cm}^{-2}$ drops by 10.6\%.
Common metal ion
species observed by COS all decline by 23-33\%, which was also found
by \citet{opp12} and shown to provide as good if not better agreement
with the available observations.  Hydrogen and metal-line absorption
primarily arise from $T<10^5$ K gas, which is reduced with PIE cooling,
and more so for metals.  The relative differences between these two
vzw simulations using CIE or PIE metal cooling is less than it would be for a
stronger wind model such as that used in \citet{wie09a}, because with
vzw winds diffuse metals are placed at higher overdensities where they
cool rapidly in either cooling scheme \citep{opp12}. 

In \S7, we compare these vzw wind models to two other wind models, a constant wind and a no wind model, both with $48 \hmpc$ size and $2\times 384^3$ resolution. The constant wind model presented here has \cite{wie09a} metal-line cooling, included because \cite{opp12} found that PIE cooling was more important for metals pushed to lower densities by these stronger winds.  We use the same no wind simulation as in \cite{opp10}. Since metals rarely reach densities where PIE metal cooling is important, there is no need to rerun a no wind simulations with PIE metal cooling.

Finally, the cosmology used in our r-series simulations is:
${\Omega}_{\rm m}=0.28$, ${\Omega}_{\Lambda}=0.72$, $h=0.7$, spectral
index $n=0.96$, ${\sigma}_{8}=0.82$, and ${\Omega}_{\rm b}=0.046$.
This agrees with the WMAP-7 constraints \citep{jar11}. The gas particle
mass is ${3.56}\times 10^{7}$ ${\rm M}_{\odot}$, and the dark matter
particle mass is $1.81\times 10^{8} {\rm M}_{\odot}$ giving an effective
galaxy mass resolution of about $2.3\times 10^9 {\rm M}_{\odot}$
and a dark matter halo mass resolution of about $1.21\times 10^{10}
{\rm M}_{\odot}$.  In this paper we will focus on galaxies and halos
generally well above these resolution limits. Unless otherwise specified, we quote distances in physical (not comoving) units, and we usually refer to them in Mpc rather than ${h}^{-1}$ Mpc.

\subsection{Generating Spectra with {\sc Specexbin}}

We use our spectral generation code {\sc Specexbin} to calculate
physical properties of the gas. {\sc Specexbin} is described in
more detail in \S2.5 of \cite{opp06} and more recently in \S2.3 of
\cite{dav10}.  Briefly, {\sc Specexbin} averages physical properties
including the gas density, the temperature, the metallicity and the
velocity of SPH particles in physical coordinates along a sight line.
It then uses look-up tables calculated with CLOUDY \citep[{}][version 08.00]{fer98} to find the ionisation fraction for the relevant ionic species.

These look-up tables, functions of density, temperature, and redshift, have
been calculated assuming collisional plus photo-ionisation equilibrium
with a uniform \citet{haa01} background. {\sc Specexbin} then converts
to velocity coordinates using Hubble's Law, accounting for the peculiar
velocities of the SPH particles, and adds thermal broadening using the
temperature and atomic weights of the various ion species. To match the
mean observed absorption
we multiply the \citet{haa01} spectrum by factors of 1.5
for the vzw model, and 1.11 for the no wind and constant wind models,
as found in \cite{dav10}.

Since our last published work with {\sc Specexbin}, we have updated the
code to include an approximate physically motivated self-shielding from
the ionisation background.  This is significant here owing to the fact
that we are interested in absorption close to galaxies that may arise
in fairly dense (and possibly self-shielded) gas.  The self-shielding
correction is purely local and is done on a particle-by-particle
basis. We compute the fraction of each particle's mass that is internally
self-shielded from the ionising background by integrating the \ion{H}{i}
column density inwards from the particle's edge, assuming that it has
a density profile given by the SPH smoothing kernel, until the column
density crosses a threshold; within that radius we assume that all
the hydrogen is neutral.  We choose a threshold of $10^{18}$cm$^{-2}$,
which provides the best fit to the neutral fractions obtained in the
full radiative transfer models of \citet{fau09}.  In practise, because
of a reasonably tight correlation between column density and physical
density~\citep[e.g.,][]{dav10}, this results in a fairly sharp density
threshold of 0.01~cm$^{-3}$ above which the gas is fully neutral.  If the
gas density is greater than $0.13$~cm$^{-3}$ our star formation algorithm
assumes that it is star-forming.  In this case we assume that the gas is
completely neutral, which is reasonable since the cold fraction in the
two-phase medium always dominates by mass over the hot fraction. This
choice of self-shielding correction also moves all the magnesium into
\ion{Mg}{ii}, which is what one expects for magnesium in neutral hydrogen based
upon the ionisation potentials of magnesium.  For random sight lines, we rarely
probe these densities, but for targeted lines of sight particularly within
10~kpc of a galaxy, the effects of self-shielding can be significant. We also put in a transition from neutral to molecular HI for ISM gas using the \cite{bli06} pressure criterion. 

Our spectral extraction implicitly assumes that all metals are in the
gas phase, but the observed correlation of quasar colours with projected
separation from foreground galaxies suggests substantial amounts of
intergalactic dust \citep{Menard10}.  \cite{Zu11} show that our vzw
simulations can reproduce the \cite{Menard10} observations if roughly
25\% of extragalactic metals (by mass) are depleted onto dust with
an SMC-like grain size distribution.  Precise accounting is difficult
because of uncertainties in the observations, in the composition
and size distribution of the dust grains, and in the destruction and
production rates as a function of time and galactocentric distance.
Dust depletion is an inevitable source of uncertainty in metal-line
absorption predictions from cosmological simulations; it could plausibly
lower our predicted absorption for refractory elements such as
carbon and silicon by 25-50\%.

\subsection{Ion Selection}

We investigate six different species: \ion{H}{i}1216, \ion{Mg}{ii}2796,
\ion{Si}{iv}1394, \ion{C}{iv}1548, \ion{O}{vi}1032, and \ion{Ne}{viii}770.
These represent some of the most common metal ions that COS probes,
along with \ion{H}{i}.  All the metal lines have doublets, making their
identification in observed spectra more straightforward.  While \ion{H}{i}
in the \lya~forest is associated with diffuse baryons tracing large-scale structures in the Universe \citep[e.g.,][]{dav99}, stronger
\ion{H}{i} absorbers are associated with higher-density structures in
the CGM~\citep[e.g.,][]{fum11,van11} and are sensitive to the outflow model
used in the simulation \citep[e.g.,][]{dav10} as well as the cooling
implementation~(see Table~\ref{table:sims}).

\hskip0.001in \ion{Mg}{ii} is one of the most common observed lines in
QAL spectra, and it appears to be associated with the CGM environments
of galaxies, especially star-forming ones \citep[e.g.,][]{kac07}, is independent of galaxy color \citep[e.g.,][]{che10},
and may indicate recent outflows \citep[e.g.,][]{bor11, bou12, mat12} and/or inflowing gas \citep{kac10}.  \ion{Mg}{ii} is often associated
with \ion{H}{i} absorbers of $N_{\ion{H}{i}}=10^{16.5}-10^{21.0} {\rm
cm}^{-2}$~\citep[e.g.,][]{hel98}, indicating that \ion{Mg}{ii} frequently
arises from self-shielded gas.  Since our cosmological simulations do
not have the resolution to fully resolve this self-shielding, we make the
assumption that all magnesium in a self-shielded region is \ion{Mg}{ii},
because the ionisation potential of \ion{Mg}{ii} is above the \ion{H}{i}
ionisation energy while \ion{Mg}{i} is below it.  We cannot resolve
the true substructure of \ion{Mg}{ii}, but we hope to understand the
typical densities and environments from which this absorption arises.
This prescription produces equivalent widths that roughly match recent
observations by \cite{rub12}, suggesting that our \ion{Mg}{ii} modelling
is not dramatically wrong.

\hskip0.001in  \ion{Si}{iv} and \ion{C}{iv} are relatively strong
lines that fall within the far-UV window often probed by COS.  These are
mid-ionisation potential lines that typically arise in optically thin gas,
but still probe fairly overdense regions at low redshifts~\citep{opp12}.

\hskip0.001in  \ion{O}{vi} is the most commonly observed high ionisation
potential metal absorber at low redshifts~\citep[e.g.,][]{tum11, dan08,tho08,bre07,sto06,che09,pro11,sem03, tri00,tri08}.  \citet{opp09} examined
the nature of diffuse \ion{O}{vi} absorbers in these simulations using
random lines of sight through their simulation volume and found that it
traces mostly photo-ionised gas in the diffuse IGM, warm-hot gas, and gas
within halos. 
Here we focus on \ion{O}{vi} absorption within halo gas,
which we will compare to the previous results from \cite{opp12} on diffuse IGM \ion{O}{vi}.

The highest ionisation potential line that we consider is \ion{Ne}{viii},
and it is valuable because it can trace $10^{5-6}$ K gas \citep{sav05}
better than any other UV resonance line.  Nonetheless, as argued in
\citet{opp12}, very weak \ion{Ne}{viii} can arise in gas photo-ionised
by the metagalactic background.  Recent observations of \ion{Ne}{viii}
by \cite{tri11} and \cite{mul09} have been interpreted to suggest that \ion{Ne}{viii} absorption arises from a transitional phase on the surfaces of cold
clouds moving through hotter material.  Unfortunately, our simulations
are not able to resolve such surfaces, so we may be missing this
component of \ion{Ne}{viii} absorption in our models.  Nonetheless,
we do find a significant number of such absorbers, in rough agreement
with data~\citep{opp12}.

\subsection{Generating and Analysing Simulated Spectra}

We examine both targeted and random lines of sight (LOS). For our targeted
LOS we randomly select central galaxies for each of three different halo mass
bins: ${\rm10}^{10.75-11.25} \msolar$ (labelled ${\rm 10}^{11} \msolar$)
; ${\rm 10}^{11.75-12.25} \msolar$ (labelled ${\rm 10}^{12} \msolar$)
and ${\rm 10}^{12.75-13.25} \msolar$ (labelled ${\rm 10}^{13} \msolar$).
For \mlo~and \mmid, we select 250 galaxies, while for \mhi~there are only
86 central galaxies in the simulation 
so we use all of them in our sample.  Because we
do not properly model the detailed internal structure of the ISM in our
galaxies, we do not present results through the centres of the galaxies,
but rather start at $b=10$ kpc.  We choose impact parameters {\it b}
ranging from 10 kpc (centre) out to 1 Mpc, with the spacing increasing with
{\it b}.  For each impact parameter {\it b}, we produce four LOS per galaxy,
meaning x+{\it b}, x-{\it b}, y+{\it b}, y-{\it b}, for a total of 1,000
LOS per {\it b} per mass bin (344 for \mhi).

 We also generate random LOS, to compare with our targeted LOS.  For our
random LOS, we choose a $100\times 100$ grid spanning the simulation in $x-y$
space, for a total of 10,000 LOS.  These are similar to the spectra
generated for \citet{opp12}.

After generating optical depths with {\sc Specexbin}, we construct
artificial spectra by convolving our data with COS's line spread function
(LSF) and adding Gaussian random noise with S/N=30 per $6 \kms$ pixel.
The LSF is roughly Gaussian with a $\approx 17$~km/s FWHM, but it has
some non-Gaussianity in the wings.  Also, there is a small wavelength
dependence to the LSF; for simplicity, we choose the G130M LSF at
1450\AA\ as representative of all wavelengths to avoid complications in
our interpretations owing to variations in the LSF.

\begin{figure*} 
\includegraphics[width=0.9\textwidth]{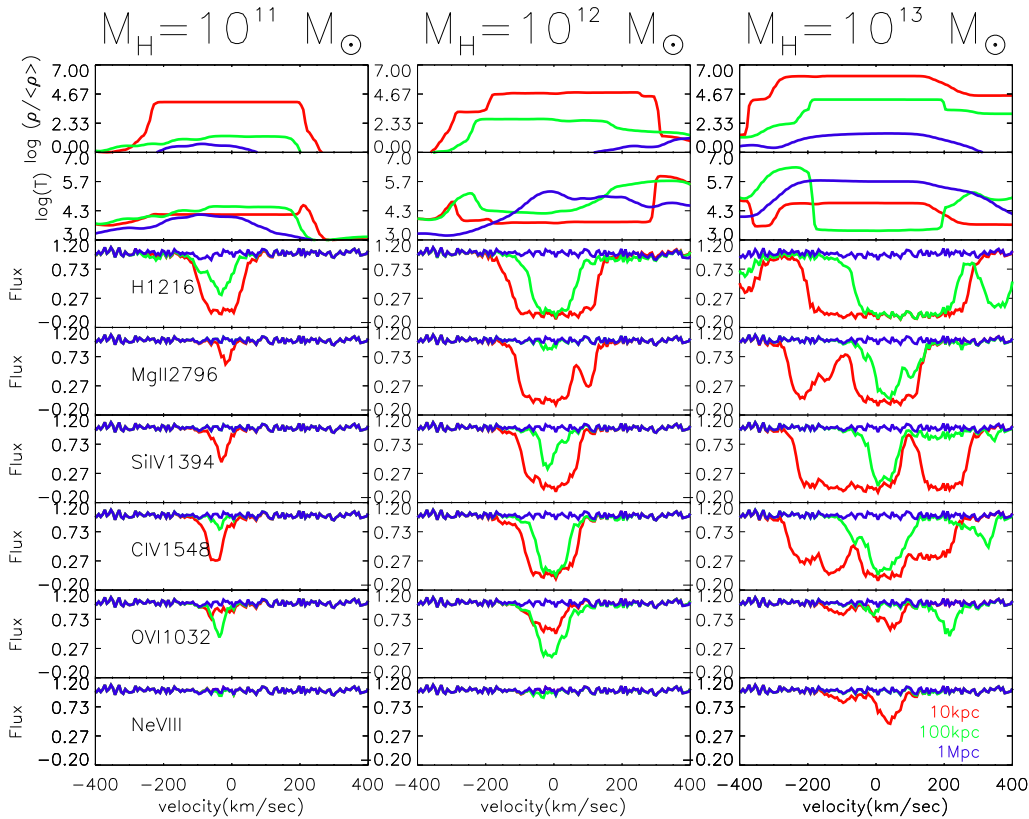}
\caption{Simulated spectra for three galaxies at z=0.25, convolved with the COS line
spread function and noise added with a S/N=30. 
Each column corresponds to a representative
  galaxy with the labelled halo mass.  We plot lines of sight at distances of
  10 kpc (red), 100 kpc (green), and 1 Mpc (blue) away from the
  galaxy. The top two panels show the \ion{H}{i} optical depth-weighted density
  and temperature, and the
  lower panels show simulated spectra for \ion{H}{i} and other
  ions. All units are physical, for a Hubble parameter $h=0.7$.}
\label{spectra}
\end{figure*}

Figure \ref{spectra} shows examples of our simulated spectra. For each
column, a single central galaxy is chosen from that mass bin, and we
plot spectra at impact parameters of 10 kpc (red), 100 kpc (green),
and 1 Mpc (blue) away from that galaxy. These particular galaxies were chosen for this figure since they have many absorption features, which illustrates the methodology well, although that is not necessarily typical.

The top two panels show the
gas temperature and density weighted by \ion{H}{i} optical depth, and
the lower panels show simulated spectra for \ion{H}{i} and other ions
as labelled.  As expected, the gas density is higher at smaller impact
parameters, which yields stronger absorption in most ions.  Also, more
massive halos generally have more absorption.  While this figure only
shows three selected galaxies, one can already see some of the trends that
will be explored and quantified later in this paper, although other trends are statistical in nature and not well depicted in this figure.

We fit Voigt profiles to the absorption features using \autovp\
\citep{dav97}, which yields column densities, line widths, (rest)
equivalent widths, and redshifts for each absorber.  We set the detection
significance criterion to $4\sigma$, which should be quite conservative
given that the noise is Gaussian random by construction.  Since the
details of deblending absorption features into multiple components can be
quite sensitive to the noise level and algorithmic details, we generally
focus on absorption {\it systems}, in which we combine all lines that
have separations $<100$~km/s. This mitigates the sensitivity to these
issues, and allows for a more robust comparison to the observations,
at the cost of discarding some of the information available about the
internal kinematics of the absorption features. Some of our figures below plot median line properties, and since weak lines outnumber strong lines, median properties necessarily depend on the lower cutoff. Our results should be taken to refer to the population of lines detectable with typical S/N=30 COS spectra, with closely separated components combined into systems.

\begin{figure*}  
 \subfigure{\setlength{\epsfxsize}{0.49\textwidth}\epsfbox{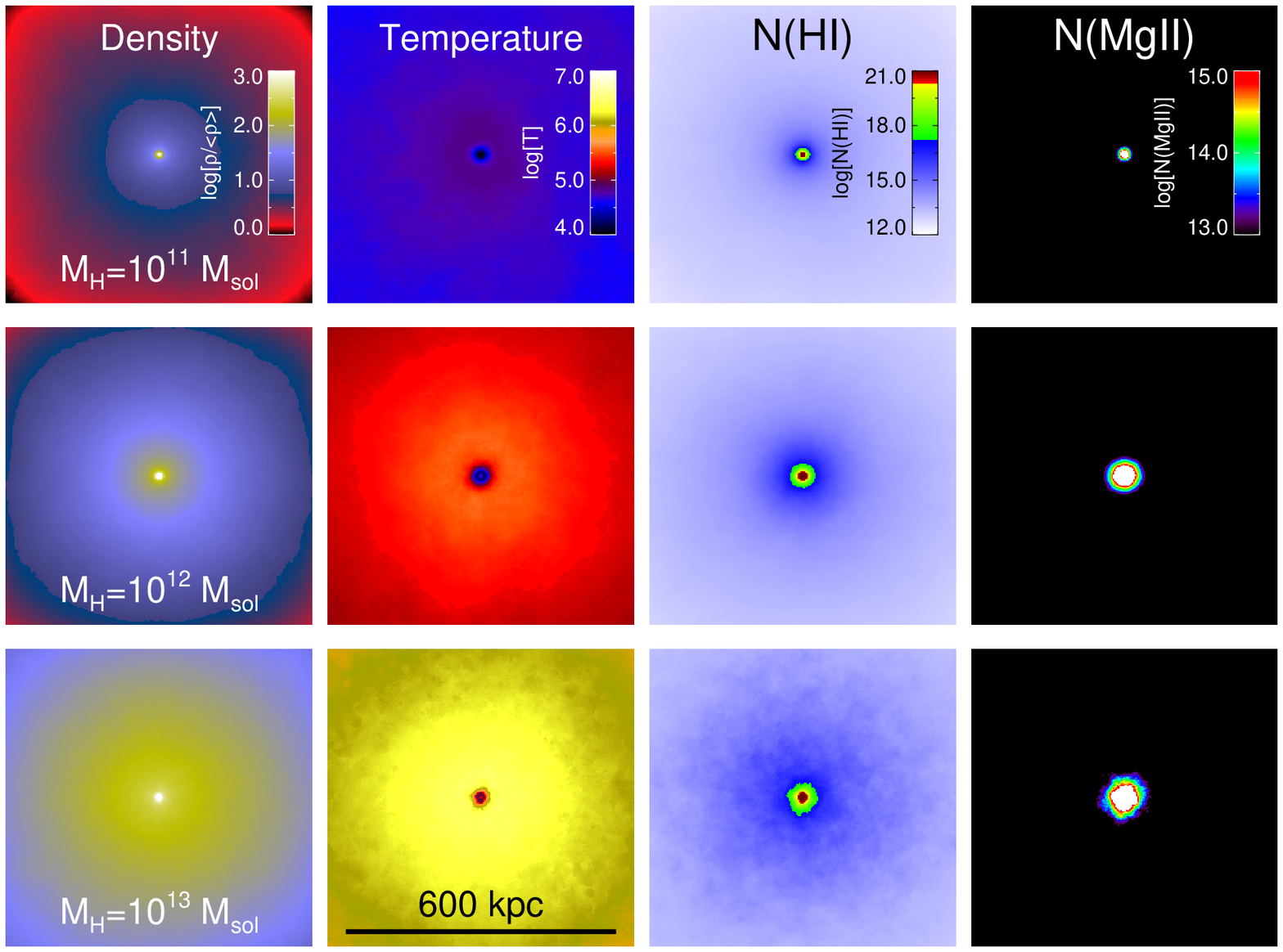}}
 \subfigure{\setlength{\epsfxsize}{0.49\textwidth}\epsfbox{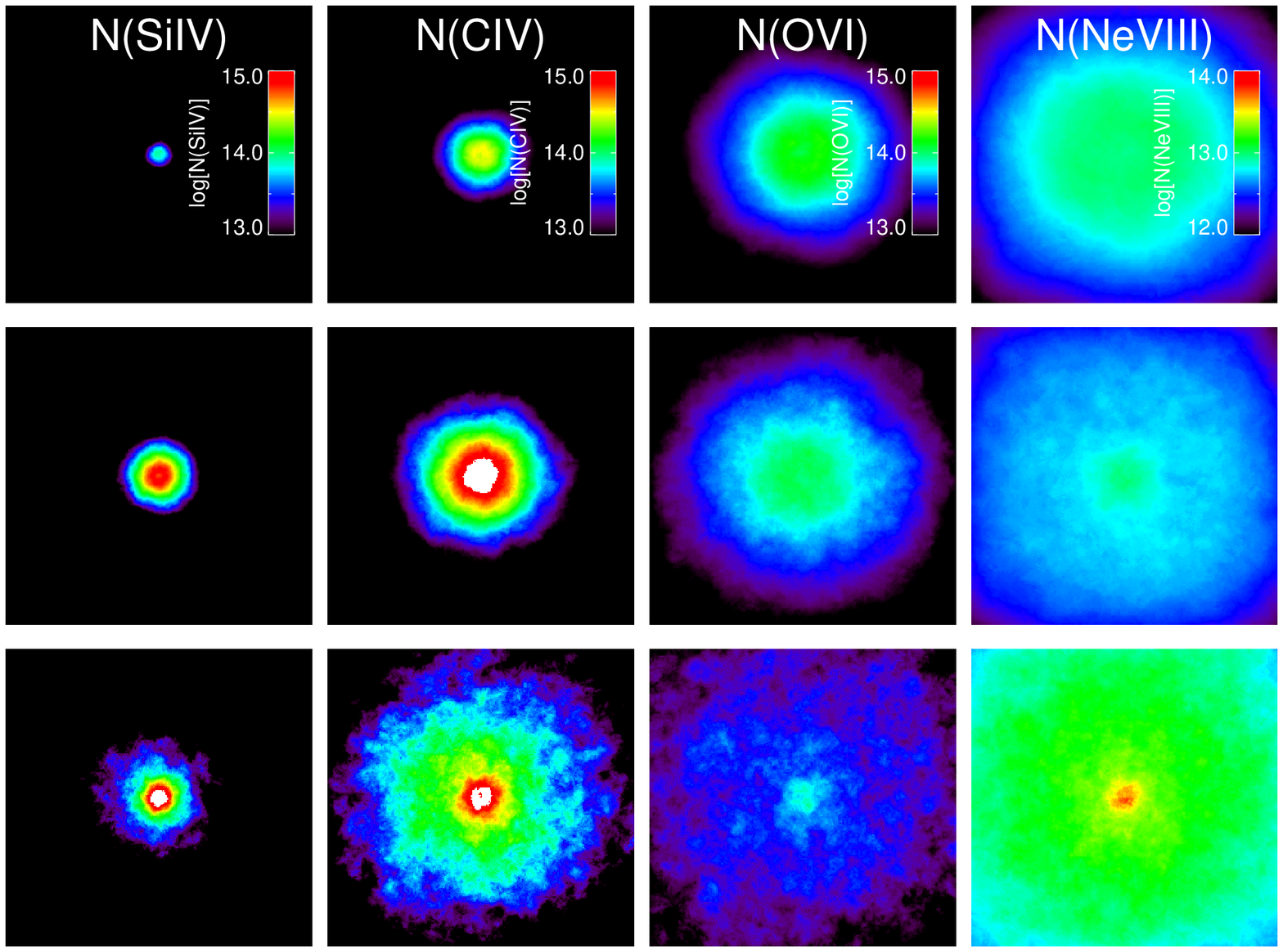}}


 \subfigure{\setlength{\epsfxsize}{0.491\textwidth}\epsfbox{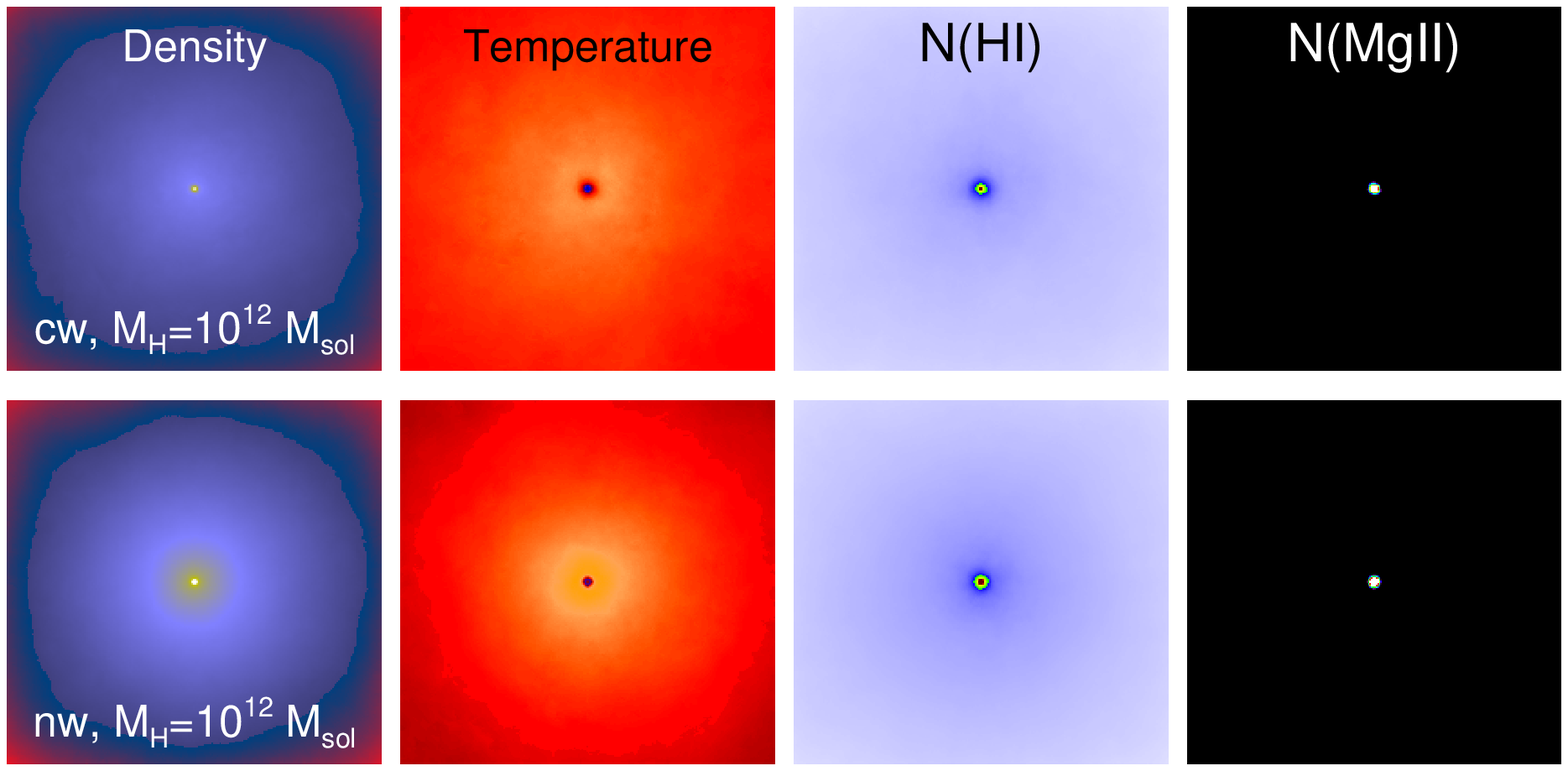}}
 \subfigure{\setlength{\epsfxsize}{0.491\textwidth}\epsfbox{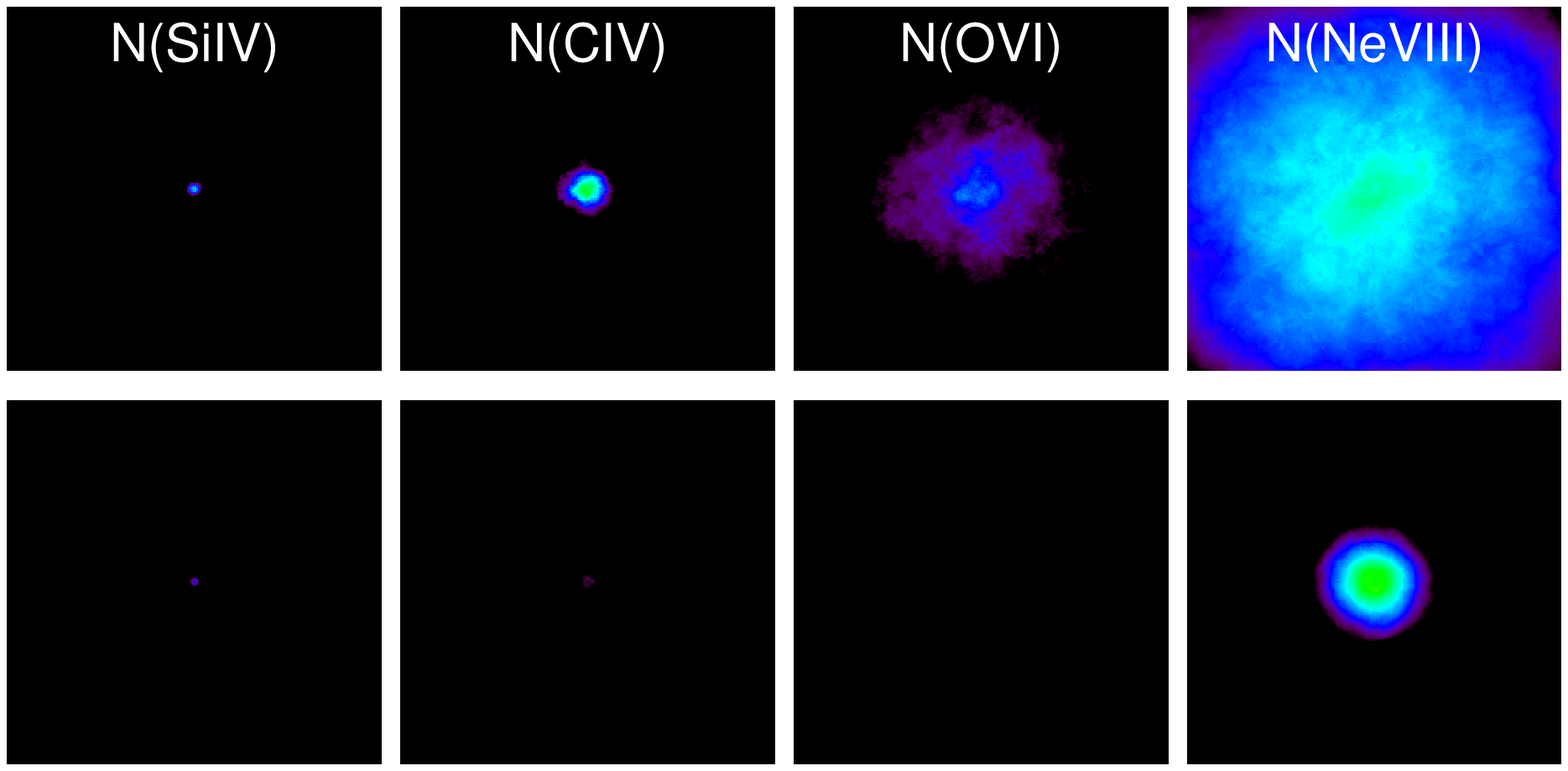}}
 \caption{Columns one and two show the median log overdensity and median log
temperature; subsequent columns show median log column densities for each ion as labelled, all at z=0.25. First three rows are for vzw; the first row is for mass bin \mlo, second for \mmid, and the third for \mhi. The fourth and fifth rows are for \mmid~halos in the constant wind and no wind models, respectively, and are discussed further in \S 7. Note the colour scales for \ion{H}{i} and \ion{Ne}{viii} are different than for the other ions. All panels are 658~kpc across.}
\label{snap300}
\end{figure*}

\section{Physical Conditions of Absorbers}
\label{sec:pca}

\subsection{The CGM in Absorption}

We seek to understand the density and temperature of the CGM and how
those conditions give rise to the presence of various observable ions.
We begin by presenting a pictorial overview of the simulated CGM, as
seen in physical conditions and line absorption, to build our intuition
as to how these physical conditions relate to observables.

Figure \ref{snap300} shows a stacked median image of the galaxies
around which we generate our targeted lines of sight.  We generate
the images by taking the pixel-by-pixel median of each given quantity
for the 250 galaxies in the mass bin of \mlo\ (top row) and \mmid\
(middle row), and the 86 galaxies in the \mhi\ mass bin (bottom
row).  Because this is a pixelized median, the satellites are
effectively removed.  The depth of the image is 658~kpc centred at
the redshift of the galaxy ($\approx 1.2$\% of the full simulation
depth).  The first two columns show the median overdensity and
median temperature of all the gas.  The remaining plots show, in
each column, the absorption in the various ions that we consider
in this paper, again for each of our three halo mass bins. Note the
colour scale for HI, subdividing \ion{H}{i} into damped ${\rm
Lyman}\alpha$ systems(red), Lyman-limits sytems (green) and the
${\rm Ly}\alpha$ forest (blue), consistent with \cite{pon08} and
\cite{voo12}, is different from the metal lines. We also include,
in the lower two panels, results for two different wind models, a
constant wind ``cw" model and a no wind ``nw" model for the \mmid\
halo mass bin (see \S7).

As expected, the gas density is strongly concentrated towards the
(stacked) central galaxy.  Galaxies in more massive halos have a
larger extent. The temperatures increase with mass bin, with the
\mhi\ mass bin showing halo gas that is predominantly near the
virial temperature, but other halos showing more sub-virial
temperatures. This reflects the now well-understood division at
\mmid\  between cold and hot-gas-dominated
halos~\citep[e.g.,][]{ker05,dek06,gab11}.  Some cool gas must be
present in \mhi\ mass halos, however, since we find significant
\ion{H}{i} absorption (see \S3.4). We reiterate that our simulations
do not include a prescription to quench star formation in hot halos,
which is required to produce red and dead galaxies~\citep{gab11},
so more realistic models might have less cool gas in \mhi\ halos.
However, recent observations (Thom et al. 2012, in preparation),
suggest that even early type galaxies have prevalent \ion{H}{i}
absorption, so perhaps whatever mechanism quenches star formation
does not greatly impact the cooler halo gas.

The morphology of the absorption (neglecting individual satellite
contributions) is strongly dependent on the ion being probed.
\ion{Mg}{ii} is highly centrally concentrated, and shows very little
qualitative morphological difference between the different halo
masses (on this plot scale).  At the opposite extreme, \ion{Ne}{viii}
shows very diffuse, extended absorption in the lower mass halos,
and is more centrally concentrated in the highest halo mass bin.
The other ions lie in between these two extremes, progressing
smoothly through \ion{Si}{iv}, \ion{C}{iv}, and \ion{O}{vi}.  We
foreshadow our upcoming results by ordering these ions by increasing
ionisation potential, which we will demonstrate provides an intuitive
guide towards understanding how absorption traces physical conditions.

\subsection{Phase Space Plots}
 \begin{figure*}
 \vskip-0.2in
 \subfigure{\setlength{\epsfxsize}{0.38\textwidth}\epsfbox{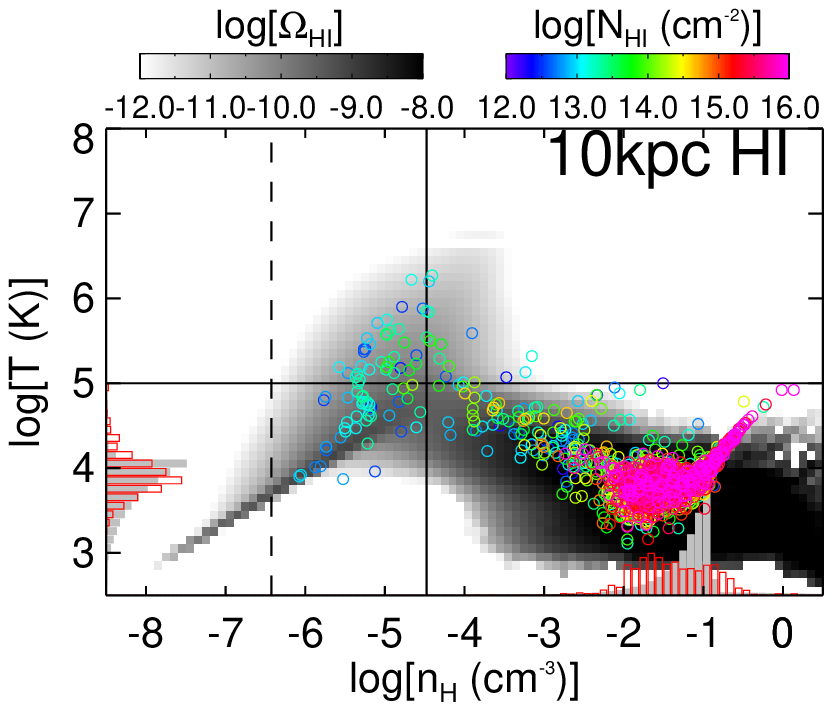}}
 \subfigure{\setlength{\epsfxsize}{0.38\textwidth}\epsfbox{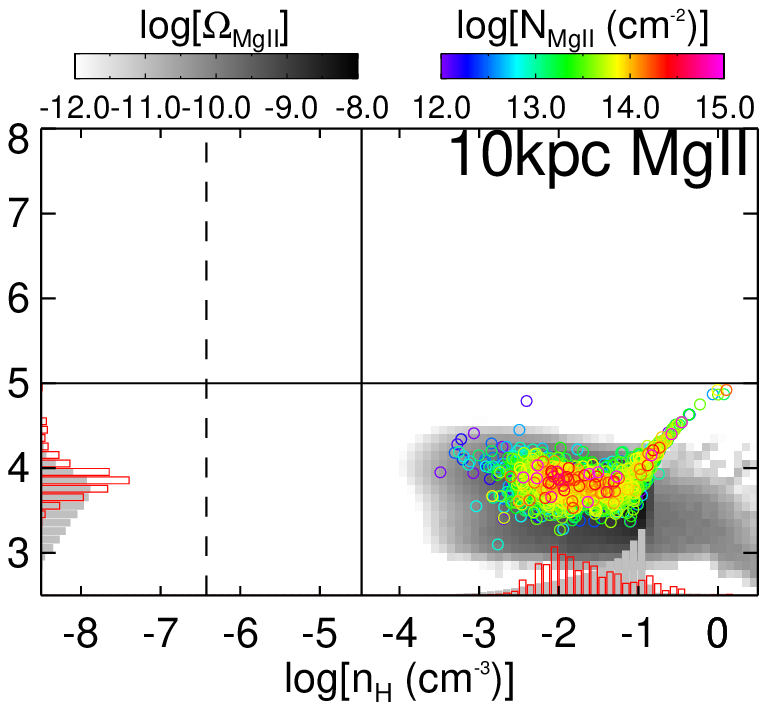}}
 \subfigure{\setlength{\epsfxsize}{0.4\textwidth}\epsfbox{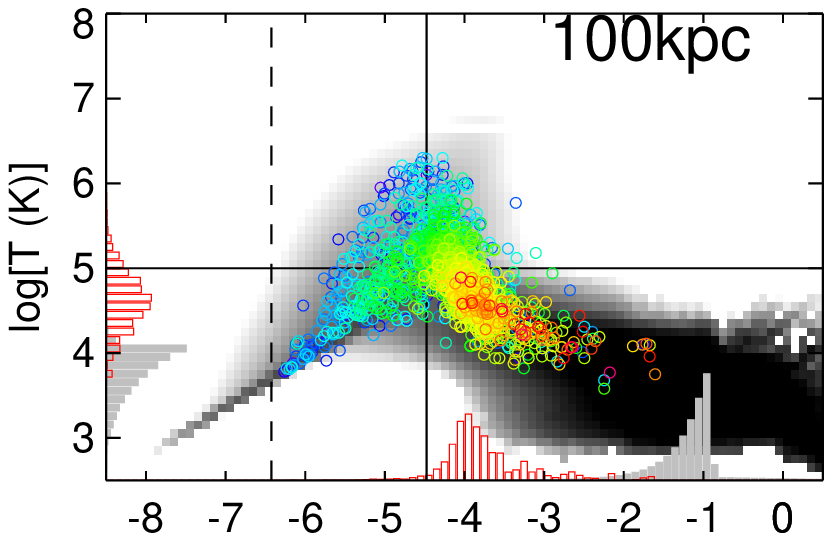}}
 \subfigure{\setlength{\epsfxsize}{0.365\textwidth}\epsfbox{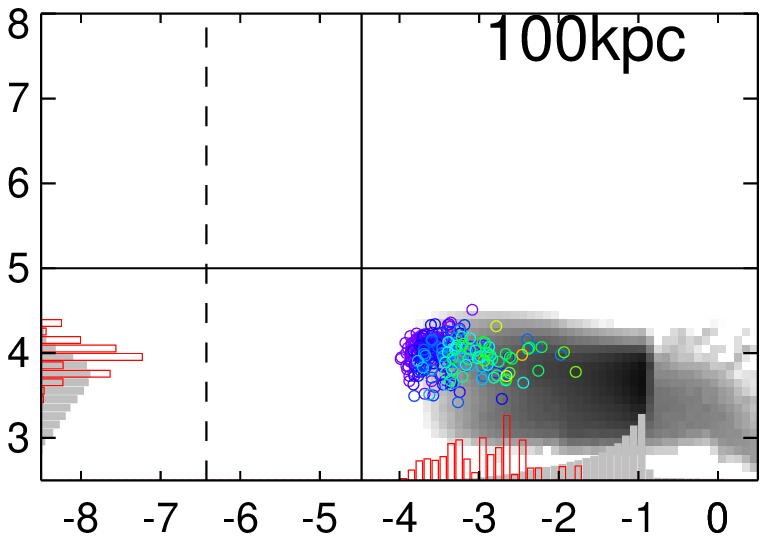}}
 \subfigure{\setlength{\epsfxsize}{0.4\textwidth}\epsfbox{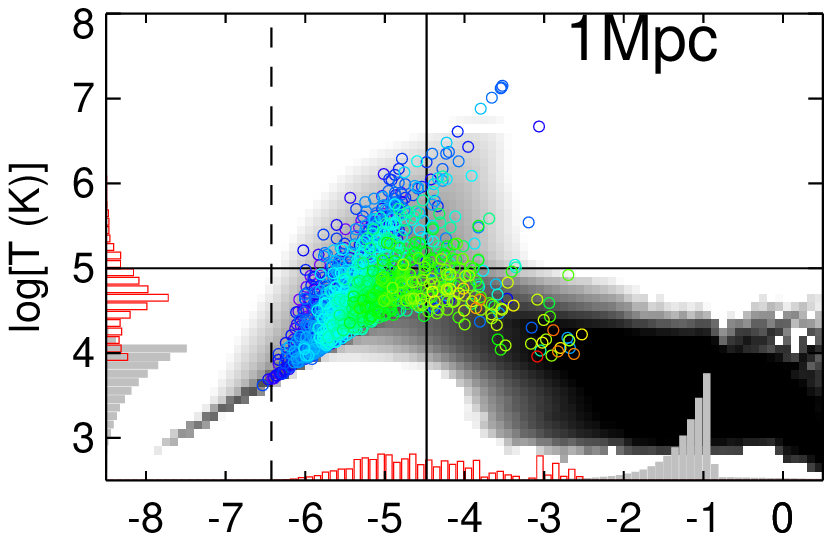}}
 \subfigure{\setlength{\epsfxsize}{0.365\textwidth}\epsfbox{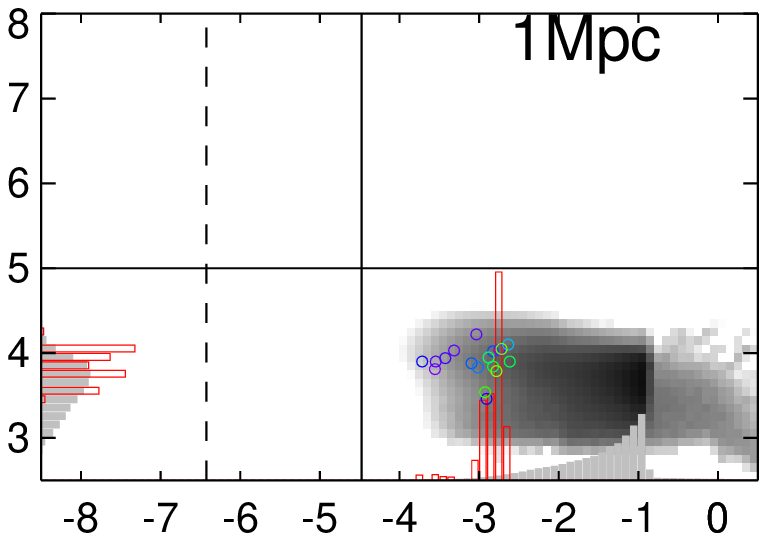}}
\subfigure{\setlength{\epsfxsize}{0.4\textwidth}\epsfbox{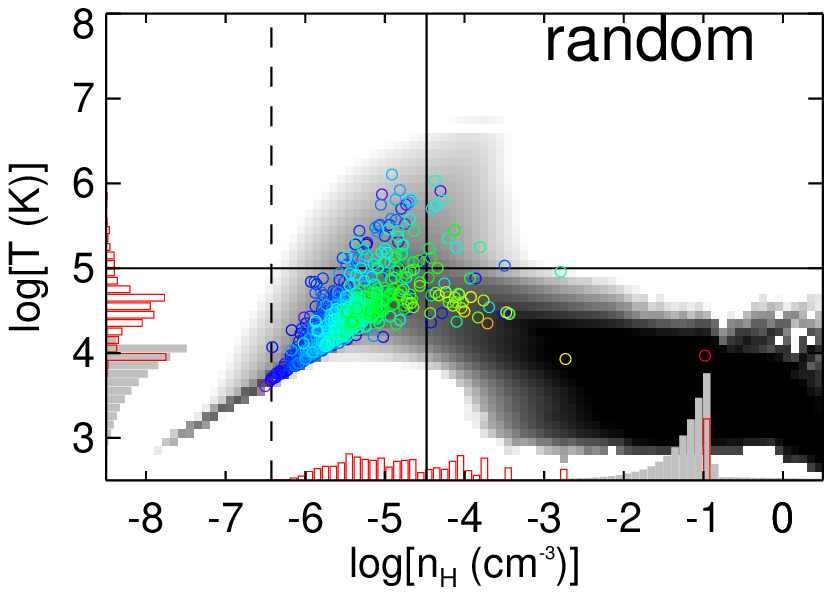}}
 \subfigure{\setlength{\epsfxsize}{0.37\textwidth}\epsfbox{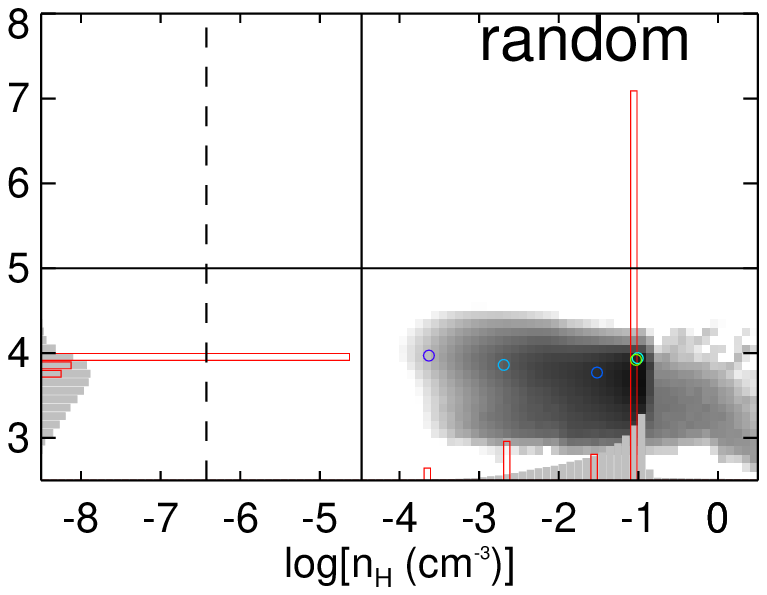}}
\caption{Phase space plots showing the location, in density and temperature
space, of \ion{H}{i} (left) and \ion{Mg}{II} (right) for ${\rm
M}_{h}=10^{12}$~${M}_\odot$ at z=0.25. Each coloured point represents an
absorption system found along targeted lines of sight at 10~kpc
(top row), 100~kpc (second), 1~Mpc (third), and along random
lines of sight (bottom).  The grey shading shows the mass-weighted absorption
for all gas below ${n}_{H}={0.13}~{\rm cm}^{-3}$ outside of all galaxies in the simulation volume, at each location in phase space. This 
is identical in each plot of a given ion.  The red and grey histograms 
show the distributions in identified absorbers and the entire volume,
respectively, collapsed along each axis. The histograms are linearly scaled, and the integral under the red and grey histograms are set to be equal. Note that the absorber colour scale is different for \ion{H}{i}. }
\label{color_low}
\end{figure*}

\begin{figure*}
 \subfigure{\setlength{\epsfxsize}{0.38\textwidth}\epsfbox{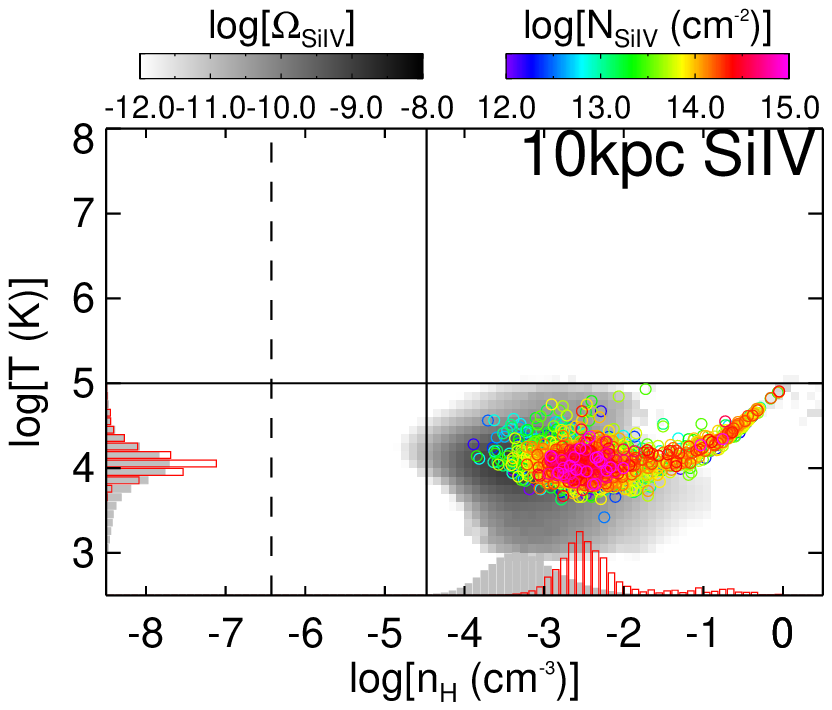}}
 \subfigure{\setlength{\epsfxsize}{0.38\textwidth}\epsfbox{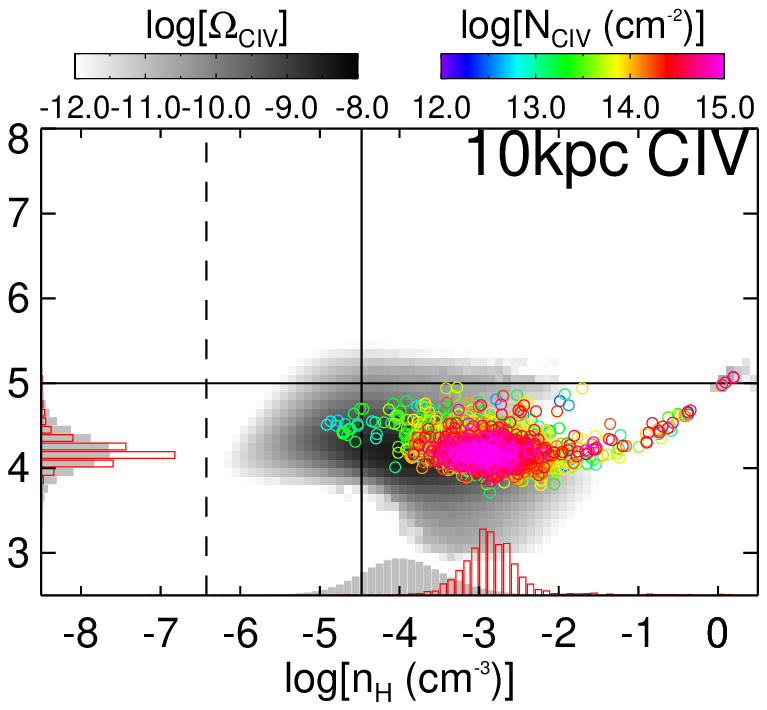}}
 \subfigure{\setlength{\epsfxsize}{0.4\textwidth}\epsfbox{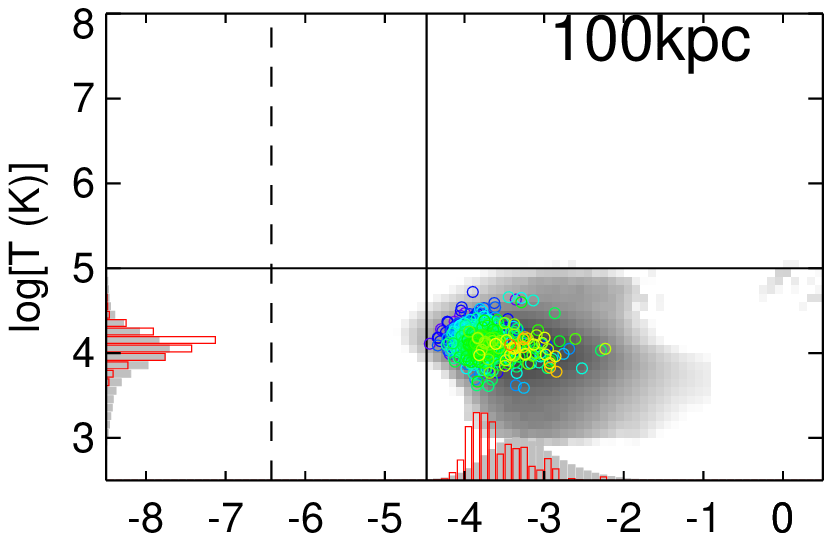}}
 \subfigure{\setlength{\epsfxsize}{0.365\textwidth}\epsfbox{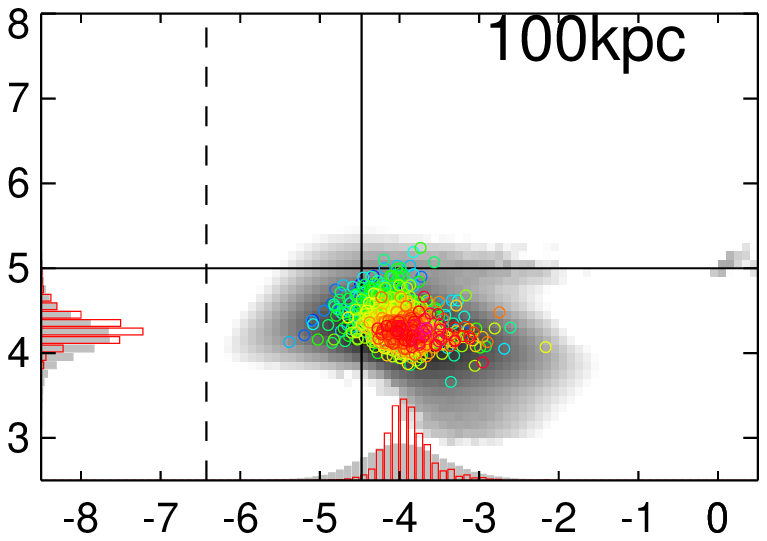}}
 \subfigure{\setlength{\epsfxsize}{0.4\textwidth}\epsfbox{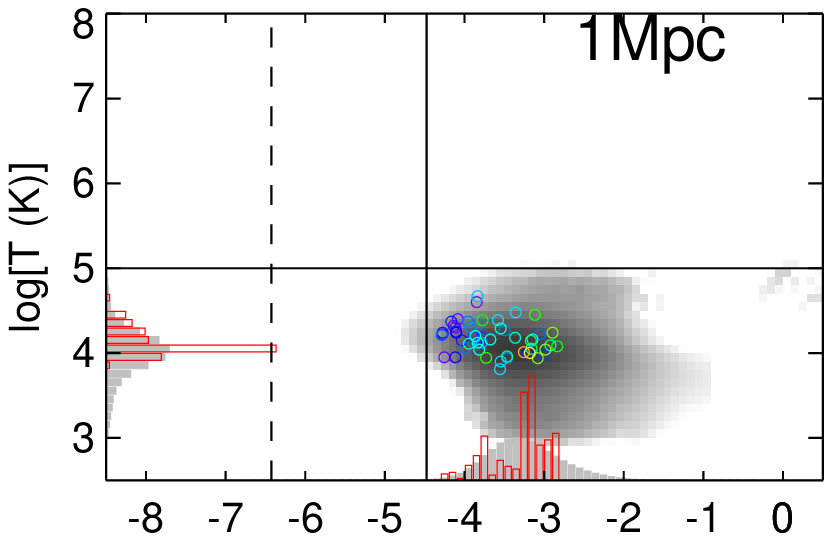}}
 \subfigure{\setlength{\epsfxsize}{0.365\textwidth}\epsfbox{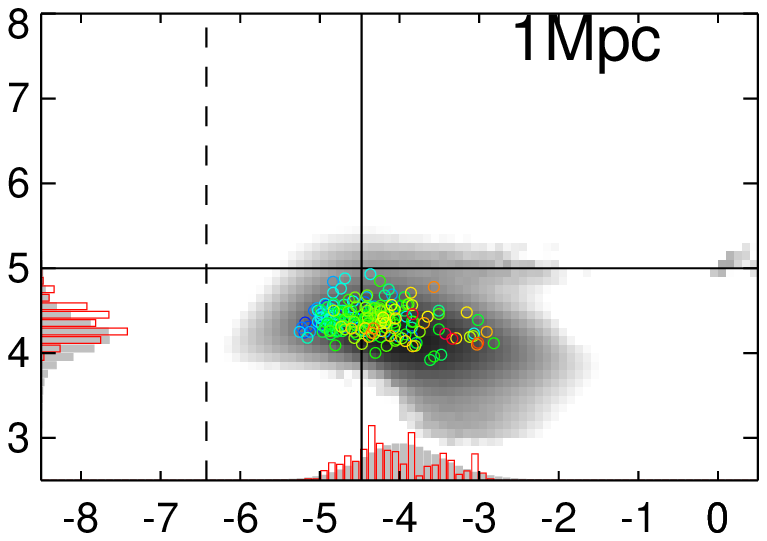}}
 \subfigure{\setlength{\epsfxsize}{0.4\textwidth}\epsfbox{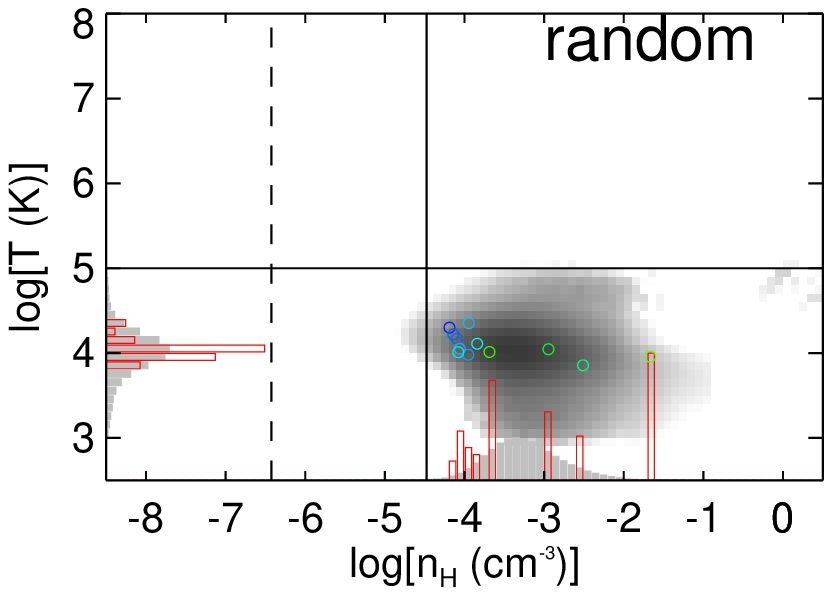}}
 \subfigure{\setlength{\epsfxsize}{0.37\textwidth}\epsfbox{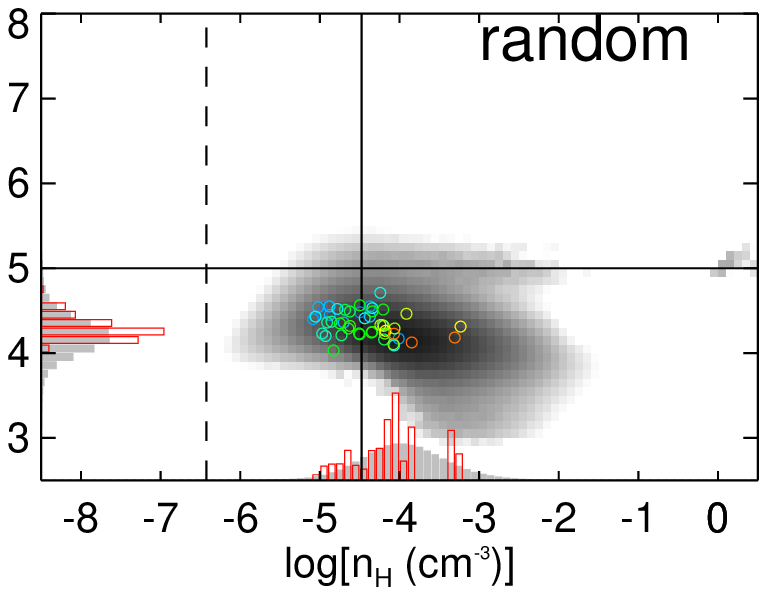}}
 \caption{Analogous to Figure~\ref{color_low}, phase space plots showing the
location, in density and temperature space, of \ion{Si}{iv} (left) and
\ion{C}{iv} (right) absorbers, for ${\rm M}_{h}=10^{12}$~${M}_\odot$,
at the impact parameters indicated.}
 \label{color_mid}
\end{figure*}

\begin{figure*}
 \subfigure{\setlength{\epsfxsize}{0.38\textwidth}\epsfbox{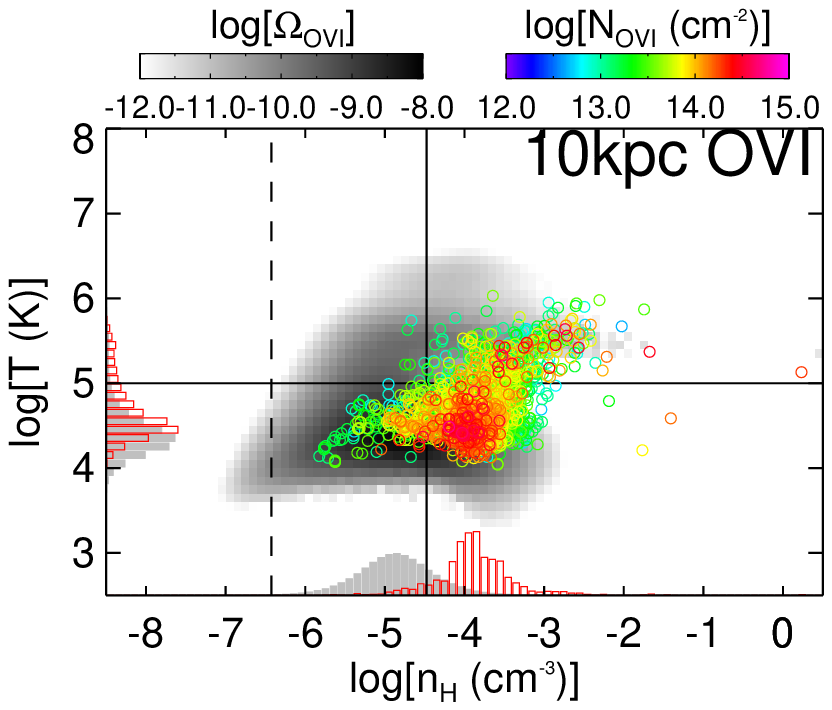}}
 \subfigure{\setlength{\epsfxsize}{0.38\textwidth}\epsfbox{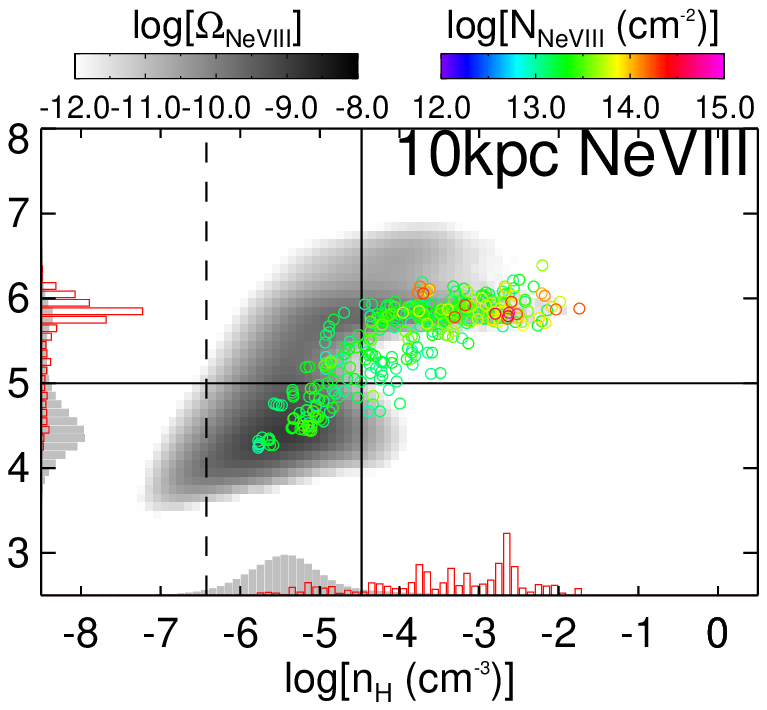}}
 \subfigure{\setlength{\epsfxsize}{0.40\textwidth}\epsfbox{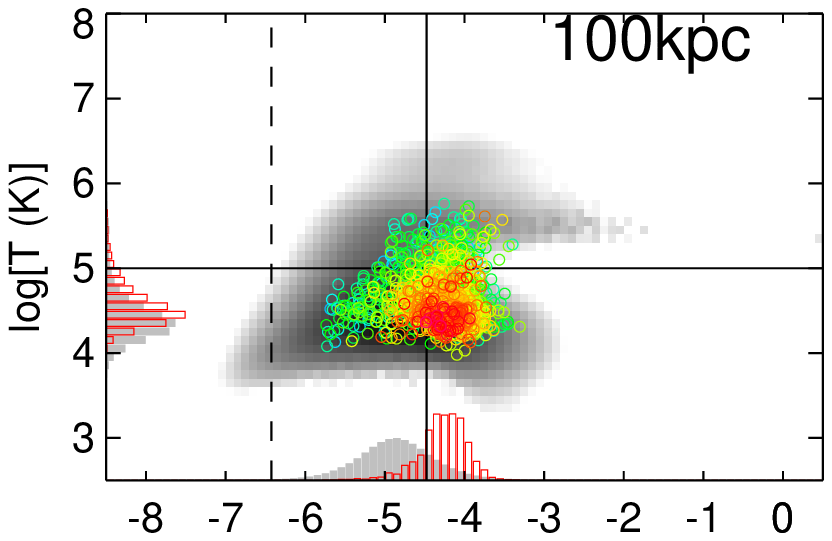}}
 \subfigure{\setlength{\epsfxsize}{0.365\textwidth}\epsfbox{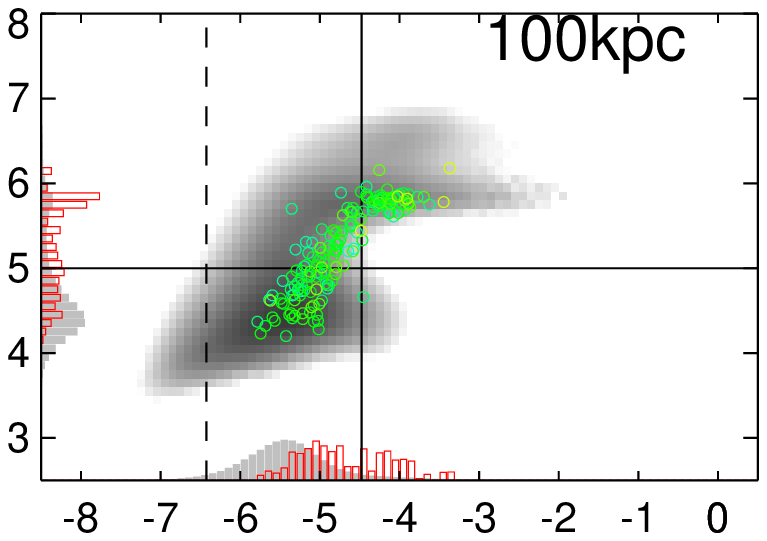}}
 \subfigure{\setlength{\epsfxsize}{0.40\textwidth}\epsfbox{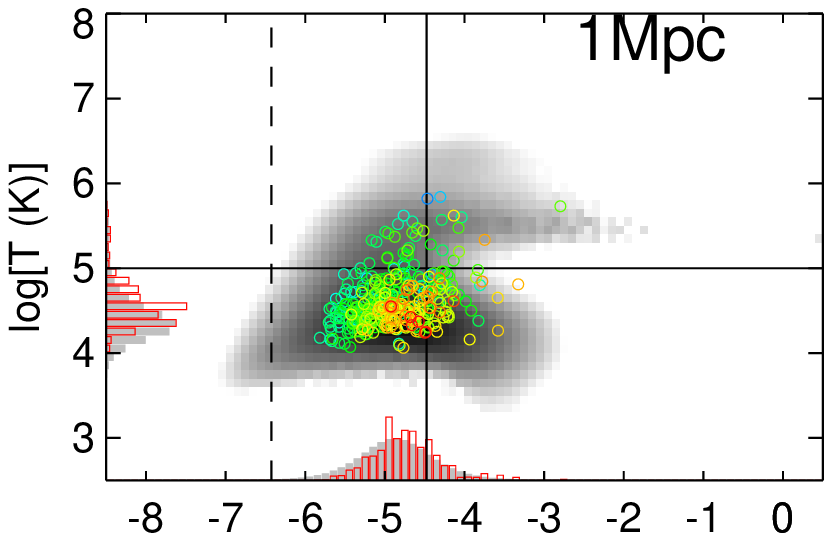}}
 \subfigure{\setlength{\epsfxsize}{0.365\textwidth}\epsfbox{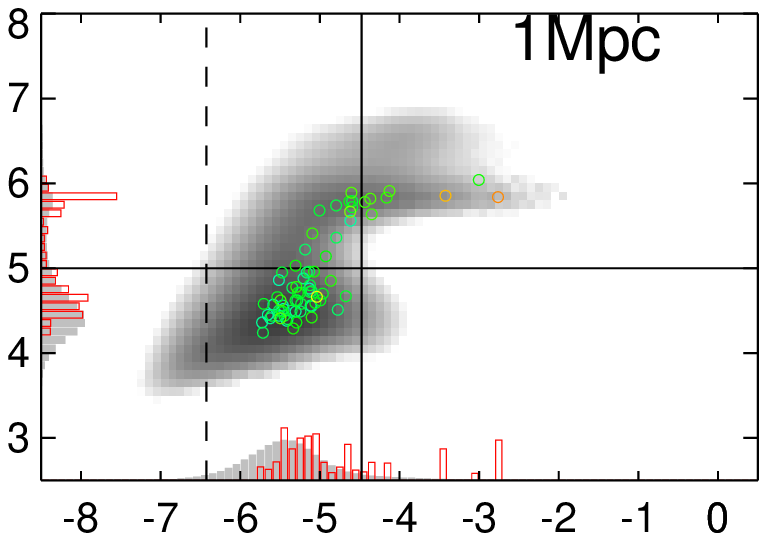}}
 \subfigure{\setlength{\epsfxsize}{0.40\textwidth}\epsfbox{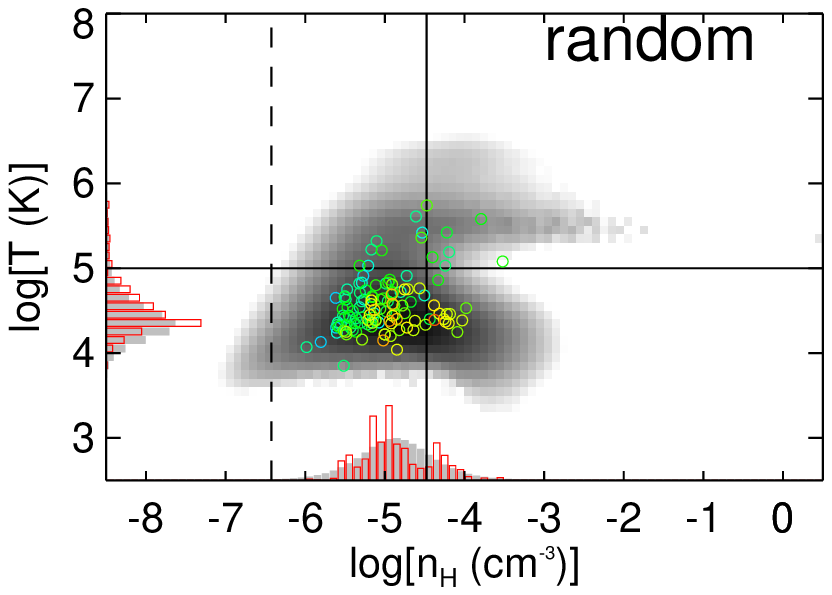}}
 \subfigure{\setlength{\epsfxsize}{0.37\textwidth}\epsfbox{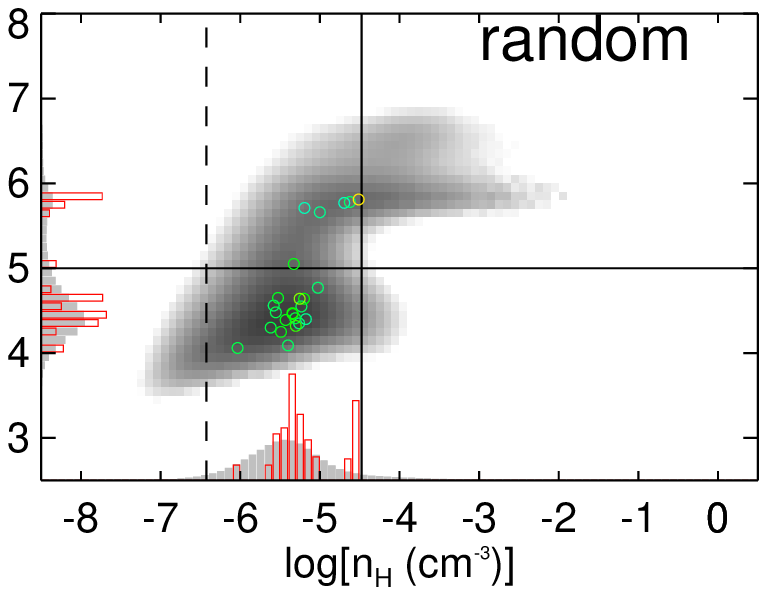}}
 \caption{Analogous to Figure~\ref{color_low}, phase space plots showing 
the location, in density and temperature space, of \ion{O}{vi} (left) and
\ion{Ne}{viii} (right) absorbers, for ${\rm M}_{h}=10^{12}$~${M}_\odot$,
at the impact parameters indicated.} 
\label{color_high} 
\end{figure*}

To begin to quantify the physical conditions giving rise to \ion{H}{i}
and metal absorption around galaxies, it is useful to examine the location
of absorption systems in density-temperature phase space.  This provides
a valuable overview of the physical conditions of the gas traced by
the various absorption species and sets the stage for discussing how
various observable properties trace these underlying physical conditions.
We thus begin by examining where our systems appear in phase space as
a function of impact parameter.

In Figures \ref{color_low}--\ref{color_high} we show the
overdensity-temperature space locations of absorption systems
(circles, colour-coded by column density) at different impact radii
as labelled, overlaid on the global (i.e. cosmic) mass distribution
of that ion ($\Omega_{\rm ion}$) computed for all non star-forming
gas (${n}_{H}<{0.13}~{\rm cm}^{-3}$) in the simulation volume (grey
shading).  We plot this for galaxies in our $10^{12}$~${M}_\odot$
halo mass bin. The vertical dashed line shows the cosmic mean density
at $z=0.25$ ($n_{\rm H}=10^{-6.43}$ cm$^{-3}$), and the solid
horizontal and vertical lines subdivide this cosmic phase space
into the four regions defined by \citet{dav10}. The temperature cut
at $T=10^5$~K divides ``hot'' phases from ``cool'' phases, and the
density cut distinguishes halo gas ($n_{\rm H}>10^{-4.47}$ cm$^{-3}$,
i.e. the virial radius density at $z=0.25$) from diffuse, intergalactic
gas.  The four quadrants, going clockwise starting at the upper
right, therefore, correspond to: hot halo, condensed, diffuse, and
WHIM gas.


In targeted sight lines, components found within $\pm 300 \kms$ of
the targeted galaxy are considered to be associated with the galaxy
(as described further in  \S\ref{sec:redshift} ). Those components
are then grouped into systems.  Our definition of a system is all
the components that lie within $100 \kms$ of any other component
within that system. We do this because fitting detailed line profiles
can be non-unique and quite sensitive to the assumed S/N, as described
further in \S2.5. The temperatures and densities of the systems are
calculated as the column density-weighted means of the individual
components.  In most cases, a single component dominates a system.

The histograms along the temperature and overdensity axes indicate
the relative fractions of the cosmic mass density (grey histograms)
and the absorption systems found in our lines of sight (red) as a
function of temperature and overdensity.  Comparing the red and
grey histograms indicates how well the absorption seen in each ion
at a given impact parameter traces the underlying density and
temperature distribution of all such ions within the volume.  Note
that the histograms are linearly (not logarithmically) scaled, and
that the integral under the red and grey histograms are set to be
equal.  The grey histograms are the same for all plots of a given
ion, and sum only gas outside galactic ISM.  In contrast, the red
histograms vary depending on the absorption found for the targeted
sight lines at various {\it b}.  To avoid outliers dominating the
computation, the red histograms have been generated with a cap in
column density of $10^{16}$ cm$^{-2}$ for \ion{H}{i} and $10^{15}$
cm$^{-2}$ for the metal lines.  All the lines above these values
have been set to this value when summing the column density-weighted
histogram, to avoid having the histograms skewed by single large,
saturated absorbers, since such absorbers generally have highly
uncertain column densities from Voigt profile fitting.  This affects
1.3\% of all \ion{H}{i} absorbers and no more than 0.7\% of all
metal-line systems.

The vertical column of panels show the distribution at three different
targeted impact parameters, 10~kpc, 100~kpc, and 1~Mpc, along
with random LOS (bottom) for each ionic species.  The different
columns correspond to different ions, beginning with \ion{H}{i} and
then ordered by increasing ionisation potential (discussed below in
Figure~\ref{lowhi}), namely \ion{Mg}{II}, \ion{Si}{iv}, \ion{C}{iv},
\ion{O}{vi}, and \ion{Ne}{viii}.  Each galaxy has four LOS per impact
parameter, and there are 250 galaxies per mass bin.  Therefore, each
panel of the top three rows represent 1000 lines of sight.  We use a
velocity window of $\pm 300 \kms$ for the targeted LOS throughout, which
we explain in \S4 as the rough window containing the majority of absorption
associated with the galaxy.  Hence, the total path length in each
panel is $6\times 10^5 \kms$ or $\delta z=2.52$.  In the bottom row,
we subsample the absorbers from the random lines of sight chosen to
cover an equivalent redshift pathlength.

The first clear trend from these figures is that the number of systems
goes down with increasing impact parameter.  The rate at which the number
drops shows some differences among the various ions; we will explore
this further in \S 6.  The straightforward explanation for this is that
both the metallicity and the gas density drop as one moves away from the
galaxy, which translates into less metal absorption.  Nonetheless, it is
clear that even at 1~Mpc, there are more systems than in the random LOS
(see \S 6).

Examining the plots more closely, one sees that for some ions there
is a distinct shift in the overdensities probed by that ion as one
moves out in impact parameter.  This is most clear for the higher
ionisation potential lines of \ion{O}{vi} and \ion{Ne}{viii}, where
the peak of the red histograms moves to lower overdensities at
higher impact parameter, but it is also true for all other ions
except \ion{Mg}{ii}.  We remind the reader that the grey histograms
are the same in every panel for each species because they correspond
to the cosmic mass density of that ion, and thus we can interpret
the red histograms as tracing a subset of the cosmic mass density.

To illustrate some of the information contained within these plots
and histograms, consider \ion{C}{iv}.  At impact parameters of
10 kpc, \ion{C}{iv} traces an average $\nh\sim 10^{-3}$ cm$^{-3}$
(red histogram), but the cosmic mass-averaged \ion{C}{iv} density
is $\nh\sim 10^{-4}$ cm$^{-3}$ (grey histogram).  In fact, there is
very little cosmic \ion{C}{iv} at $\nh>10^{-3}$ cm$^{-3}$.  However,
10 kpc from a galaxy inside a $10^{12}$ M$_\odot$ halo corresponds to
a rare location in the Universe where there are high levels of metal
enrichment and high densities, and hence \ion{C}{iv} traces these metals.
At impact parameters of 100 kpc, the red histogram for \ion{C}{iv}
peaks at $10^{-4}$ cm$^{-3}$, similar to the cosmic averaged peak in
the grey histogram.  However, the distribution of densities is narrower,
indicating that absorbers at this impact parameter do not account for the
full range of densities giving rise to all the \ion{C}{iv} absorption.
At impact parameters of 1 Mpc, \ion{C}{iv} traces overdensities similar
to the cosmic average.  Finally, the random LOS \ion{C}{iv} should
theoretically trace the cosmic distribution of \ion{C}{iv} and the red and
grey histograms should overlap.  However, as discussed in \citet{opp12},
the absorber histogram is biased towards lower densities.  The random
LOS sample covering $\delta z=2.5$ does not adequately sample densities
with $\nh$ $>$ ${\rm 10}^{-4}$ cm$^{-3}$ because this pathlength is not
long enough to probe these rare high density regions.  This illustrates
a central tenet of this paper, namely that close-in targeted LOS probe
regions of density that are not well-sampled via randomly chosen quasar
absorption sight lines.

We can similarly examine other metal ions.  Random LOS showing
absorption from \ion{Mg}{ii} usually probe gas within roughly 10~kpc of
galaxies. As a result, random LOS with \ion{Mg}{ii} absorption have the
same overdensity as the 10~kpc gas. Meanwhile, \ion{Mg}{ii} absorption
from further away generally probes lower density gas. For mid-ions, the
random LOS have significant absorption farther away, and hence 10~kpc LOS
pick out especially dense gas, but there is little distinction between
100~kpc, 1~Mpc, and random.  For high ions, these arise in more diffuse
gas so that the random LOS can probe quite low overdensities, and hence
the LOS at 10 and 100 kpc pick out particularly denser gas.

Finally, for comparison, we examine neutral hydrogen.  \ion{H}{i}
most closely resembles the general behaviours of \ion{Mg}{ii}, in
that the cosmic density of \ion{H}{i} is heavily weighted toward
$\nh > 10^{-2}$ cm$^{-3}$ and most cosmic \ion{H}{i} arises at small
impact parameters going through ISM gas.  There are, of course, a
few key differences between the well-understood behaviour of
\ion{H}{i} and a metal-line species like \ion{Mg}{ii}.  Most cosmic
\ion{H}{i} is in damped \lya~systems (DLAs), gas that is either
within or near the galactic ISM \citep{wol05}, as indicated by the
steadily rising grey histograms for \ion{H}{i}.  The \lya~forest
tracing diffuse gas, i.e. column densities $N_{\ion{H}{i}}\sim
10^{13}-10^{15}$ cm$^{-2}$, corresponds to an insignificant fraction
of the total cosmic \ion{H}{i}, unlike the high ionisation potential
metal lines where most of the cosmic absorption arises from IGM and
CGM densities.  Also, \ion{H}{i} can trace a relatively large range
of phase space because: (i) hydrogen is present at all densities,
(ii) \lya~has a high oscillator strength, and (iii) its presence
does not rely on metal enrichment.  A comparison of the red absorber
histograms to the grey cosmic density histograms makes little sense
for \ion{H}{i} since all absorbers $>10^{16}$ cm$^{-2}$ have been
reduced in our analysis to $10^{16}$ cm$^{-2}$, which hides most
of the \lya~absorption from damped \lya~and Lyman-limit systems.
However, the red histograms show a significant downward shift in
overdensity going from impact parameter of 10 kpc out to 1 Mpc,
reflecting the very different densities probed at these three impact
parameters.

\subsection{Physical Conditions vs. Ionisation Potential}

Overall, the phase space plots illustrate a key general trend: the higher
the ionisation potential of the metal species, the lower the overdensity
that it traces on average, {\it and hence the further away from galaxies
it arises}.  We will argue that this is the case in our simulation because
these metal species that we explore are primarily photo-ionised, except
for high ionisation potential lines in massive halos.

\begin{figure*}  
\includegraphics[width=0.9\textwidth]{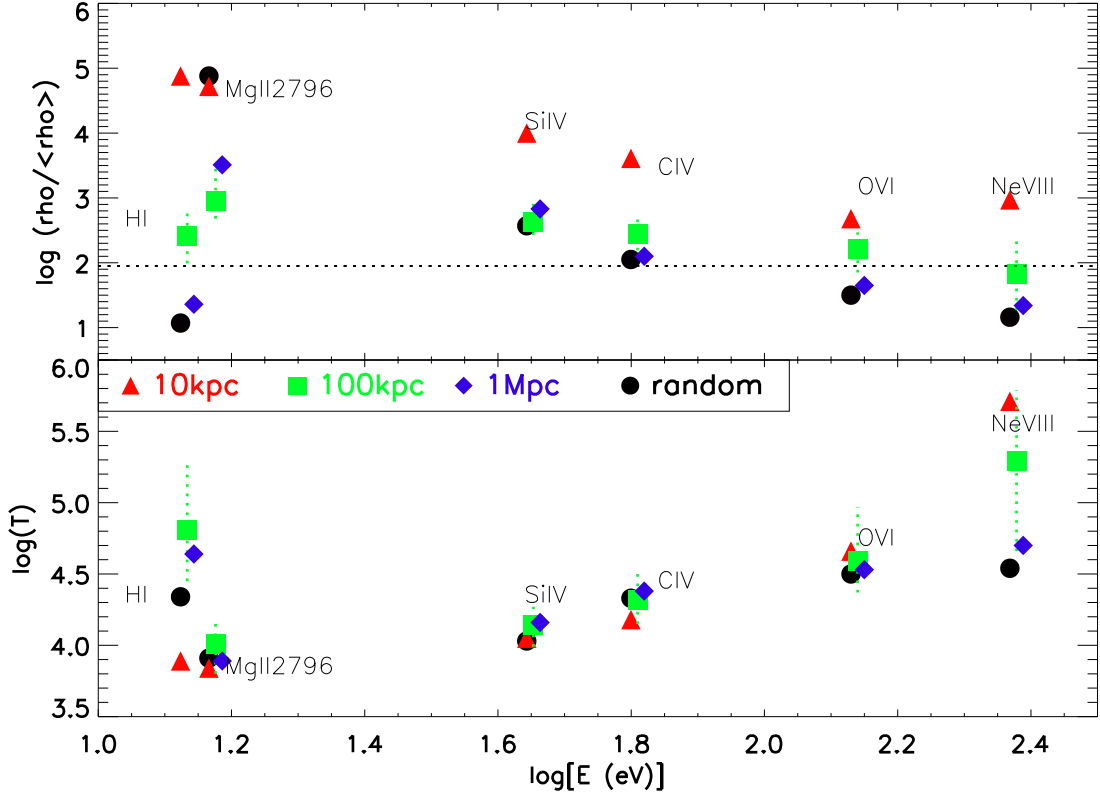}
\caption{{\it Top panel:} Median column density-weighted
  overdensity as a function of ionisation energy for the six ions we
  consider. Four points for each ion are plotted, corresponding to three
  different {\it b} values (all for mass bin ${10}^{12}$) and random
  LOS. 100 kpc (green squares) and random (black circles) are plotted at the
 energy in eV at which the labeled ion is ionised to the next higher level; 10 kpc (red triangles) and 1 Mpc (blue diamonds) are shifted for visibility. Points are plotted at the well-defined median; vertical range (not error) bars (shown only on the 100~kpc case for
  clarity) span the 16\% to 84\% enclosing values.  The dotted horizontal 
  line represents the virial overdensity at $z=0.25$.
  {\it Bottom panel:} Similar to the top panel, but for column 
  density-weighted temperatures. }
 \label{lowhi}
\end{figure*}

To quantify these trends, Figure~\ref{lowhi} plots the column
density-weighted median overdensity (upper panel) --  i.e., the
overdensity above which 50\% of the absorption occurs -- of absorption
systems of various ions, as a function of their ionisation potential
for going from that ion into the next higher ionisation state.  To
compute the column-density weighted median value, we take, for each
ion, all the systems for all the lines of sight and bin them according
to their corresponding overdensity (or temperature), as in the red
histograms from figures \ref{color_low}, \ref{color_mid} and
\ref{color_high}. We then sum the column densities of all the systems
in a given bin.  We plot the overdensity (or temperature) corresponding
to where half of the total column density (for all the bins) is
above/below that value.

We also plot the column density-weighted median temperature (lower
panel), i.e. the temperature above which 50\% of the absorption
occurs.  Points are plotted for the well-defined median for a sample
of 250 galaxies. To illustrate the range, we show vertical range
bars (not error bars) that span the 16\% to 84\% enclosing values
for the 100~kpc bin, as an example. This is the range of the
overdensity (or temperature) above which 16\% of the absorption
occurs to the overdensity (or temperature) above which 84\% of the
absorption occurs.  The different colour points represent the three
different impact parameters, with black points showing the global
values from the random LOS for comparison.  These plots are for
galaxies in halos with $M_{\rm halo}\approx 10^{12}M_\odot$. We
plot here the full range of column densities shown in Figures
\ref{color_low}-\ref{color_high}, but these findings do not depend
significantly on the column density range probed.

Figure~\ref{lowhi} (top panel) shows that the median density of
metal absorption decreases steadily with ionisation potential, but
that the rate of the decrease depends on impact parameter.  Far
from galaxies, the decrease is roughly two orders of magnitude from
\ion{Mg}{ii} to \ion{Ne}{viii}.  This reflects the predominantly
photo-ionised nature of our absorption lines arising in diffuse gas
far from galaxies, as elements require lower densities to achieve
higher ionisation levels.  The mean overdensity at 1~Mpc is essentially
indistinguishable from that of the random LOS.  At 10~kpc there is
nearly as large a decrease in the median overdensity as a function
of ionisation parameter as at 1~Mpc, but in each case the median
overdensity is around two orders of magnitude larger at 10~kpc than
at 1~Mpc.  Impact parameters of 100~kpc are an interesting counterpoint,
showing less than an order of magnitude difference in the mean
overdensity from \ion{Mg}{ii} to \ion{Ne}{viii}.  As we will discuss,
this arises because in \mmid~halos the high ionisation potential
lines start to have an additional contribution from collisionally
ionised gas that occurs in the denser regions closer to galaxies.

The horizontal dotted line approximately delineates the overdensity
boundary of virialised halos using Equation~1 from \cite{dav10}.
At 10~kpc, unsurprisingly, all the absorption in all the ions arises
in gas that is within halos. At larger impact parameters and for
the random LOS, the median overdensity depends strongly on the
ionisation state, with ions having ionisation potentials lower than
\ion{C}{iv} generally arising in halo gas for \mmid~halos, while
higher ionisation potential absorption often arises in gas with
overdensities lower than that corresponding to galaxy halos.
Unfortunately, this absorption is usually weak and
photo-ionised~\citep{opp12}, and hence generally still does not
trace the so-called missing baryons in the WHIM.

Turning to the temperatures, Figure~\ref{lowhi} shows that the
median temperature of metal absorption rises steadily with ionisation
potential. It also shows that, with the exception of \ion{Ne}{viii},
there is essentially no dependence of metal absorption gas temperature
on the impact parameter.  This reflects the fact that the majority
of metal ions all arise in $\sim 10^{4-4.5}$~K gas, i.e. photo-ionised
gas temperatures, regardless of the impact parameter.  Note that
the virial temperature in this halo mass range is about a million
degrees, the upper boundary of the plot.

For \ion{Ne}{viii} at large impact parameters, the median temperature
is around 30,000~K, still considerably lower than its collisional
ionisation peak temperature of $10^6$~K, and owes to photo-ionised
gas.  This begins to change as one approaches galaxies.  At 100~kpc,
about half of the absorption comes from gas above 200,000~K, and
at 10~kpc most of the \ion{Ne}{viii} is hot, with half the absorption
coming from gas above 500,000~K that is collisionally ionised.
Since (as is evident from the phase space plots) these \ion{Ne}{viii}
absorbers are also stronger, we obtain a result matching that of
\citet{opp12} that strong \ion{Ne}{viii} absorbers are more often
collisionally ionised.  By using our targeted LOS, we show here
that this mostly arises from \ion{Ne}{viii} within 100~kpc of
galaxies, at least in these \mmid~halos.  As hypothesised in
\citet{opp12}, these strong lines are probably the only ones that
have been studied carefully with COS to date~\citep[e.g.,][]{tri11},
but larger and deeper samples should uncover a population of
photo-ionised \ion{Ne}{viii} arising in more diffuse gas.

This hot gas close to galaxies is also evident in the \ion{H}{i}, where at
100~kpc half the \ion{H}{i} absorption traces gas that is above 50,000~K,
and much of this gas is above the halo virial density.  Even at 1~Mpc
there is some hot \ion{H}{i}, as the median temperature is still fairly
high at 40,000~K, even though the median overdensity is quite low. As
one can see from Figure \ref{color_low}, there are a number of absorbers
than can arise from both hot halo gas and truly diffuse WHIM gas.  Hence,
in principle, broad \ion{H}{i} absorbers could trace truly diffuse WHIM
gas (i.e. the missing baryons), but one has to be careful not to count
hot halo gas, which may be detectable by other means.

Our simulations predict that even high ions are mostly photo-ionised
at all radii.  This is at odds with recent work by \cite{sti12},
whose simulations find that \ion{O}{vi} is predominantly collisionally
ionised in their individual galaxy simulations.  Their model has
the advantage of having higher resolution than ours, but our models
have the advantage that our feedback prescription has been carefully
constrained to match a broad range of observations, including IGM
enrichment.  The main difference is likely that their feedback model
relies on super-heating gas within the ISM that drives hot gaseous
outflows, whereas our model follows scalings expected for momentum-driven
winds in which the gas is pushed out via radiation pressure, and
therefore is not super-heated.  Our outflows do heat once they
interact with surrounding gas, but since they are typically quite
enriched, the metal-enhanced cooling rates are rapid.  Simply put,
their feedback model adds hot gas to the halo by construction, while
ours adds cooler gas by construction; it is not immediately evident
which is closer to correct, and likely depends on the details of
how winds are actually launched which is currently not well understood.
\citet{sti12} did not show the \ion{O}{vi} temperature as a function
of radius, area-weighted as would be appropriate for an absorption
line survey, so it is difficult to compare our results directly.
But these differences highlight that modeling the CGM is not a fully
solved problem.  Examining such detailed statistics as line ratios
and alignment statistics between \ion{O}{vi} and \ion{H}{i} (and
low metal ions, e.g. \ion{Si}{iii}) provides a way to characterize
the temperature of the \ion{O}{vi} gas and discriminate between
such scenarios.  We note that at least in the random lines of sight
examined in \citet{opp09}, the observed alignment statistics between
\ion{O}{vi} and \ion{H}{i} were better reproduced in a momentum-driven
wind scalings model as opposed to our constant wind model that
yielded more collisionally-ionised \ion{O}{vi}.  We are conducting
such comparisons now against COS-Halos data.

In summary, different metal ions probe different physical conditions
around galaxies, with lower ionisation potential lines probing
denser gas.  \ion{Mg}{ii} absorbers probe very high density gas in
and around the ISM of galaxies, while high ionisation potential
lines probe diffuse gas in the outskirts of halos and beyond the
virial radius.  The absorbing gas temperatures generally reflect
the photo-ionised nature of metal absorption even down to small
impact parameters in our models, with the notable exception of
\ion{Ne}{viii} lines that can arise in hot gas near galaxies.  In
the next section we will show that \ion{O}{vi} in larger halos where
a hot CGM is present is also mostly collisionally ionised.  These
results demonstrate that spanning a range of ions with low to high
ionisation potentials can in principle probe a wide range of physical
conditions in the CGM, but that when a range of ions are seen in a
single system, it is probably unwise to assume that they all arise
from the same gas (for the purposes, e.g., of CLOUDY modeling).
Throughout the rest of this paper, we will discuss our results in
terms of low (\ion{Mg}{ii}), mid (\ion{Si}{iv} and \ion{C}{iv}),
and high ionisation (\ion{O}{vi} and \ion{Ne}{viii}) lines, since
this provides an underlying physical context for understanding the
behaviour of these various metal absorbers.

\subsection{Physical Conditions vs. Halo Mass}

The previous section focused entirely on \mmid~halos, as representative
of a typical $L^*$ galaxy halo.  However, the increasing collisional
ionisation contribution to the higher ionisation potential species
might suggest that there could be some halo mass dependence since,
as is evident in Figure~\ref{snap300} and has been shown by e.g.
\citet{ker05}, large halos have substantially more hot gas.

\begin{figure*} 
 \subfigure{\setlength{\epsfxsize}{0.79\textwidth}\epsfbox{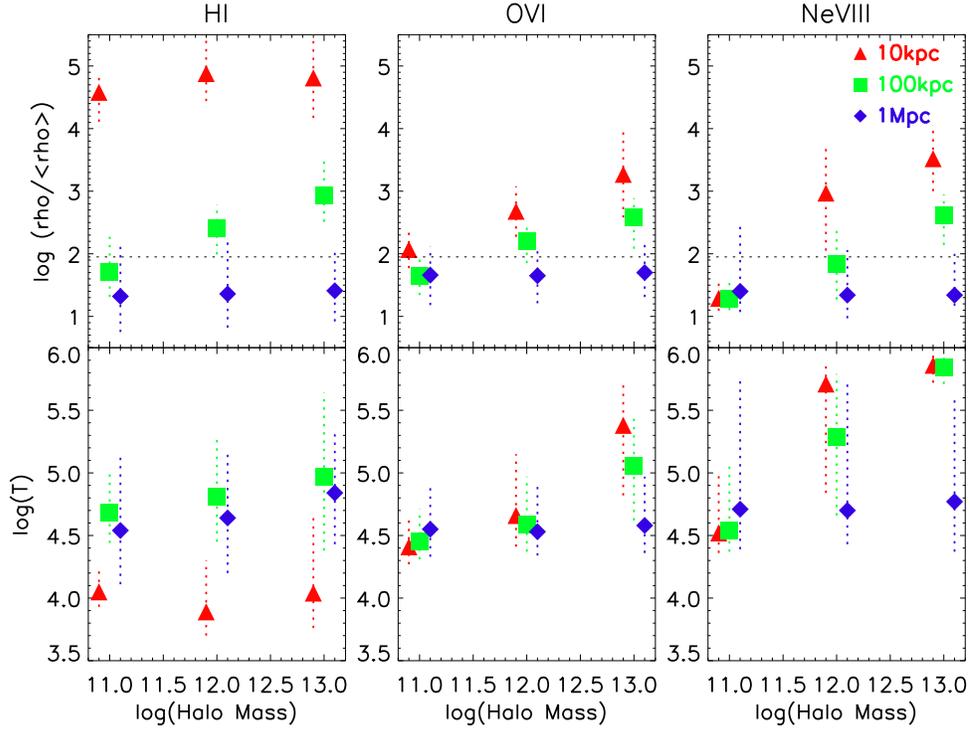}} 
\caption{Median column density-weighted overdensity and
temperature vs. halo mass for \ion{H}{i} (left panels), \ion{O}{vi} (middle panels), and \ion{Ne}{viii}
(right panels).  For all panels, the different colours correspond to the three
different impact parameters: 10 kpc (red triangle), 100 kpc (green square), and 1 Mpc (blue diamond)
(offset for clarity).  
The range bars span the 16\% to 84\% enclosing values, and the dotted
horizontal line represents the virial overdensity.}
\label{lowhi_hi}
\end{figure*}

To investigate any trends with halo mass in Figure \ref{lowhi_hi},
we plot in the top panels the column density-weighted median
overdensity and temperature as in Figure \ref{lowhi}, here as a
function of halo mass, focusing on \ion{H}{i} (left panel), \ion{O}{vi}
(middle panel), and \ion{Ne}{viii} (right panel). As in Figure
\ref{lowhi}, we plot the full range of column densities shown in
Figures \ref{color_low}-\ref{color_high}, but these findings do not
depend much on column density range probed.  We show three different
impact parameters, slightly offset horizontally for ease of visibility,
and here we show the $16-84\%$ range for all cases.  The points for
galaxies in halos of \mmid~are identical to those shown in
Figure~\ref{lowhi}.  We do not show the random LOS here; the values
are similar to the 1~Mpc case.  We also do not show the other ions,
because in those cases there are no discernible trends with halo
mass.

At 10~kpc, for \ion{H}{i}, the absorption arises from low temperature
gas for all mass bins. While Figure \ref{snap300} shows halos of
mass \mhi~have high median temperatures, Figure~\ref{lowhi_hi} shows
the HI absorption mostly comes from low temperature gas present in these
halos.  For absorption around galaxies in halos of \mlo, most of
the \ion{O}{vi} and \ion{Ne}{viii} absorption owes to gas with
overdensities lower than that found in galaxy halos, since the
densities are below the dotted virial overdensity line.  The exception
is \ion{O}{vi} at 10 kpc, where about half the absorption comes
from the outskirts of halos.  Both high ionisation potential lines
have a median temperature within the photo-ionised regime at all
impact parameters (although for \ion{Ne}{viii} there is a tail to
higher temperatures for impact parameters of 1 Mpc).  Hence \ion{O}{vi}
and \ion{Ne}{viii} absorption around galaxies with halo masses of
\mlo~actually traces photo-ionised, diffuse IGM gas. This occurs
because there is so little hot gas in these small halos~\citep{ker05}
that there is little opportunity for these ions to trace collisionally
ionised gas.

For galaxies in \mmid~halos, the temperatures are still mostly
representative of photo-ionised gas for \ion{O}{vi}, but \ion{Ne}{viii}
begins to trace somewhat hotter gas, already indicating a contribution
from collisionally ionised gas at smaller impact parameters. At
impact parameters of 10 kpc, this gas is within galaxy halos for
both ions, and even at 100 kpc almost all the \ion{O}{vi} absorption
and about half the \ion{Ne}{viii} absorption owes to gas with an
overdensity consistent with being within galaxy halos. At impact
parameters of 1~Mpc, the gas giving rise to this high ionisation
potential absorption still mostly arises from gas outside of galaxy
halos.

Finally, for galaxies in \mhi~halos, we see a strong dependence of
physical conditions on impact parameter, with the density and
temperature both being substantially higher at smaller impact
parameters.  At impact parameters of 10 kpc and 100 kpc for both
ions, all the absorption owes to gas within galaxy halos that is
at temperatures indicative of collisional ionisation.  This reflects
the substantial presence of hot, virial-temperature gas within these
halos.  At impact parameters of 1 Mpc, however, the absorption still
mostly arises from photo-ionised, diffuse IGM gas.

In summary, the high ionisation potential lines tend to trace
collisionally ionised hot gas when it is present.  But this gas is
generally only present abundantly well inside of halos, at impact
parameters of less than 100~kpc, and only in massive halos where a
hot gaseous atmosphere can form.  In those cases, these (and other
high ion) lines may trace hot gas in halos that are not easily
probed by X-ray emission (or absorption lines), offering a unique
opportunity to study these baryons.

\section{Absorption around galaxies in redshift space}
\label{sec:redshift}

The distance of an absorber from a galaxy can be observationally
characterised by two parameters: the line-of-sight velocity difference
$\Delta v$ and the impact parameter $b$.  In this section we discuss the
former, i.e. how absorption properties vary with the velocity distance
from the central galaxy.  We seek to answer questions like the following:
Does the absorption drop off with $\Delta v$ at different rates for
different ions?  Can we identify a characteristic LOS velocity distance
over which the galaxy provides a clear excess of absorption?  How do
these tendencies reflect the physical conditions of the absorbing gas?

\begin{figure*}  
 \subfigure{\setlength{\epsfxsize}{0.79\textwidth}\epsfbox{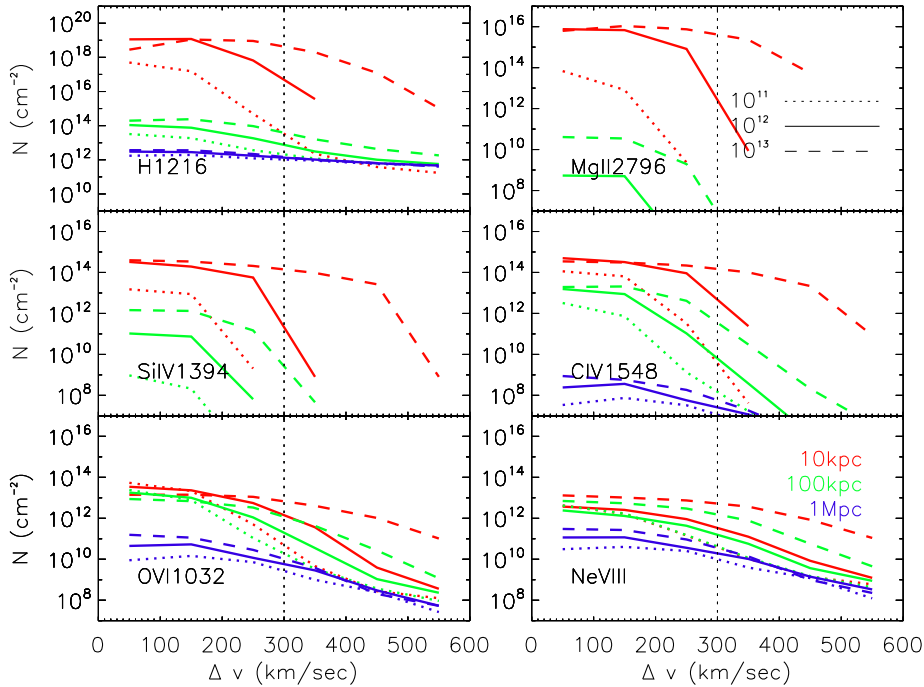}}
 \caption{Median column density (N) vs. velocity separation ($\Delta v$) 
from the central galaxy $\Delta v$ for $M_{h}={\rm 10}^{11,12,13}M_\odot$
(dotted, solid, and dashed lines, respectively).  We show results at
three impact parameters: 10~kpc (red), 100~kpc (green), and
1 Mpc (blue). Vertical dotted lines mark 300$\kms$.}
\label{ewvel}
\end{figure*}

Figure \ref{ewvel} shows the column density (N) versus velocity
separation ($\Delta v$) from the central galaxy for each of our six
ions, with the species ordered by increasing ionisation potential going
down the columns.  The lines show the median column densities for all
lines of sight at the three different impact parameters: 10 kpc (red),
100 kpc (green), 1 Mpc (blue), for the three different halo masses:
${10}^{11} M_\odot$ (dotted), ${10}^{12} M_\odot$ (solid), and ${10}^{13}
M_\odot$ (dashed). We use $100 \kms$ bins and plot at the bin's midpoint.
For example, $\Delta v=150 \kms$ plots the total column density between
$\pm(100-200) \kms$.  Note that the y axis has a rather large range;
most of this range is not accessible observationally,
but we include it to accentuate the trends.

The median column density decreases with increasing $\Delta v$ for
all the ions, following the trend highlighted before that the
absorption tends to be higher near galaxies.  Except at high halo
masses, there are drops in these curves before $\pm 300 \kms$.
Conservatively, we may say that the vast majority of absorption
generally occurs within $\pm 300 \kms$ of the central galaxy's
velocity, although in many cases the majority occurs within a smaller
velocity interval. We choose to define the velocity window associated
with the galaxy as 300 $\kms$ to account for as much absorption in
the higher mass halos as possible while not going far beyond the
drops in the lower mass halos. We note that $\pm 300 \kms$ is roughly
consistent with recent observational work by \cite{pro11}.
Quantitatively, we can consider the total amount of absorption
within $\pm 300 \kms$ relative to total absorption within $\pm 600
\kms$. Looking at all three impact parameters, \ion{O}{vi} has at
least 96\% of its column density within this velocity limit for
$M_{\rm halo}=10^{11} \msolar$, 97\% for $M_{\rm halo}=10^{12}
\msolar$, and 88\% for $M_{\rm halo}=10^{13} \msolar$. The absorption
of lower ionisation potential metal species falls off even faster
with increasing velocity difference.  This indicates that in general
one has to look only within the central 600$\kms$ window around a
galaxy to find most of the absorption associated with it.

Absorption around galaxies in our most massive halo bin, ${10}^{13}
M_\odot$, shows a shallower decrease with increasing $\Delta v$,
reflecting these halos' larger virial velocities and hence larger
peculiar motions of gas and satellites.  While the $\pm 300 \kms$
cut misses some absorption in the large halos at small {\it b}, LOS
passing so close to such massive galaxies are rare.  Henceforth,
we only consider absorption within $\pm 300 \kms$ of galaxies for
our targeted LOS in the subsequent figures.

Now we examine the differences between the various metal ions.
The low ionisation potential ions decrease in absorption more rapidly
with increasing impact parameter than high ionisation potential
ions.  The column densities of \ion{Si}{iv} and \ion{C}{iv} at fixed
$\Delta v$ drop rapidly from 10~kpc to 100~kpc, and at 1~Mpc they
are not even visible in this plot.  In contrast, the high ionisation
potential ions, \ion{O}{vi} and \ion{Ne}{viii}, show essentially
no decrease in absorption at fixed $\Delta v$ when the impact
parameter increases from 10~kpc to 100~kpc, and the drop from 100~kpc
to 1~Mpc is more modest than for low ionisation absorbers.  The
decrease in column density is also more dramatic with increasing
$\Delta v$ for the low ionisation potential ions.  This occurs
because high ionisation potential lines arise in more extended gas
distributions at lower overdensities, which trace the general
large-scale structure in which galaxies live, while the low ionisation
potential ions are more confined to the high-density gas found
closer to galaxies (as seen in Figure~\ref{snap300}).

In this plot, \ion{H}{i} behaves much like a high ionisation potential
line -- note the similarity in the shapes of the curves between
\ion{H}{i} and \ion{O}{vi}, although there is a large difference
in the magnitude of the column densities.  This reflects the fact
that \ion{H}{i} can arise in a wide range of physical conditions,
and even at small impact parameters there is a substantial contribution
to the column density from gas along the LOS out to large $\Delta
v$'s \citep{kol03,kol06}. At 1~Mpc, \ion{H}{i} absorption does not
depend much on the central halo mass.  \ion{H}{i} is observed almost
everywhere that metal ions are seen, and in almost every case has
a column density greater than any metal ion.

Finally, we examine the trends with halo mass.  Nominally, the
larger peculiar velocities within larger halos should result in
greater $\Delta v$'s; each factor of 10 in halo mass should correspond
to a factor of $10^{1/3}\approx 2.2$ in $\Delta v$. In practise,
one would expect $\Delta v$ differences that are somewhat lower
than this because one integrates the column density through the
entire halo.  Low ionisation potential ions like \ion{Mg}{ii} are
close to this expectation, with the differences between the curves
being roughly a factor of 1.7 in $\Delta v$.  This indicates that
\ion{Mg}{ii} basically arises only when the LOS intercepts a galaxy,
and at large $\Delta v$ the satellite galaxies giving rise to the
absorption are tracing the underlying halo potential.  The mid
ionisation potential ions \ion{Si}{iv} and \ion{C}{iv} are similar
to \ion{Mg}{ii}, indicating that the denser gas giving rise to these
ions also traces the underlying potential.  Moving towards higher
ionisation potential lines, we see somewhat smaller differences as
a function of halo mass, as absorption in these ions starts to pick
up gas that is outside the virial radius and hence not dominated
by the halo potential.  It is important to note once again that all
the massive galaxies in the vzw simulation have star formation rates
well above comparable-mass galaxies in the real Universe, owing to
a lack of a quenching mechanism in our simulations ~\citep[e.g.][]{gab12}.  Therefore, the
absorption trends at higher mass may not be reflective of the real
Universe, where the observed \ion{O}{vi} declines in strength in
more massive halos \citep{tum11}.  Nonetheless, to first order for
all ions, close to galaxies it is the dynamics of the host halo
that establishes the $\Delta v$ distribution of the absorption.

In summary, all ions have the vast majority of their absorption
arising within a redshift-space distance of (conservatively) $\Delta
v\pm$ $300 \kms$ around galaxies for all but the most massive halos.
Low ionisation species show a sharp drop with $\Delta v$, while
high ionisation species show a more gradual drop.  More massive
halos show a broader absorption distribution in $\Delta v$ reflecting
their larger potential wells.  For low ionisation lines, absorption
drops very rapidly with impact parameter, while for high ionisation
lines, the drop with impact parameter is much slower.  These trends
broadly reflect the overall morphology of absorption relative to
galaxies, in the sense that lower ionisation lines are more confined
to dense gas closer to galaxies.

\section{Absorption around galaxies versus impact parameter}
\label{sec:b}

\begin{figure*}
\subfigure{\setlength{\epsfxsize}{0.79\textwidth}\epsfbox{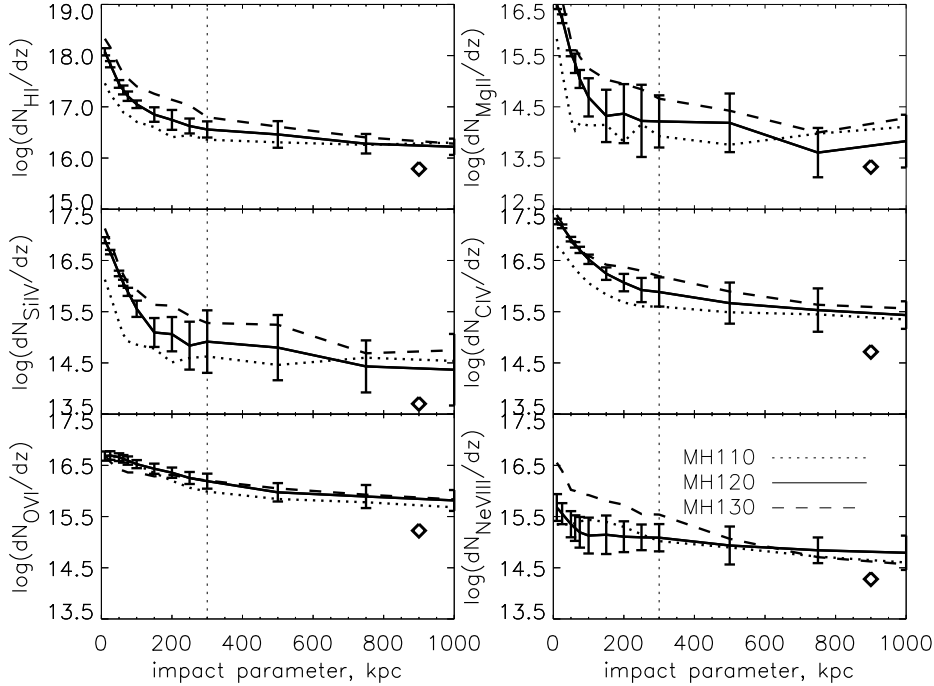}}
\caption{The total column density (not number density) per unit redshift versus
impact parameter around galaxies in halos
of $10^{11} M_\odot$ (dotted), $10^{12} M_\odot$ (solid)
and $10^{13} M_\odot$ (dashed). For halos, dN/dz is the summed column density for the given ion along all lines of sight at the given impact parameter over the velocity range $\pm 300 \kms$, divided by $\Delta z= {\rm (Number~of~halos~in~sample)}\times(600\kms)/(1+z)$. The black symbols indicate dN/dz for random lines of sight. Note that the vertical range for each ion varies,
although it always spans 4 dex. 
The vertical dotted line corresponds to 300 kpc. Error bars show the cosmic variance across sixteen simulation sections of equal volume.}
\label{NvsB}
\end{figure*}

We now turn to examining the extent of absorption around galaxies
in physical space, as quantified by the impact parameter $b$.  From
the previous section, we know that much of the absorption occurs
within approximately $\pm 300$~km/s of a galaxy (the mild exception
being for low ionisation potential lines in massive halos).  Hence,
we will examine how absorption within this $\Delta v$ range varies
with impact parameter, as a function of both ionisation level and
halo mass.

In Figure \ref{NvsB} we plot the summed column density per unit
redshift along all lines of sight with a given impact parameter,
with the impact parameters ranging from $10-1000$~kpc.  We show
results from our vzw simulation for galaxies in halos with masses
of $10^{11} M_\odot$ (dotted lines), $10^{12} M_\odot$ (solid), and
$10^{13} M_\odot$ (dashed).  We show error bars owing to cosmic
variance (since statistical errors are small). To calculate those,
we divide our simulation volume into sixteen sections of equal
volume and compute the dN/dz values for lines of sight within each
section, and compute the dispersion over the sixteen sections. We
ignore the (rare) sections that have no absorption when calculating
the dispersion.

Here, we have chosen to plot a summed column density, instead of a
median one.  This is because we are interested in predicting the
total absorption along the line of sight as one moves out in impact
parameter.  The median value can be highly dependent on the resolution
and noise level of the spectra, since better quality spectra will
result in many more weak lines.  In contrast, the summed absorption
is a more robust quantity.  However, it does have the disadvantage
that it can be biased by a single, very large absorber.  To mitigate
this, we apply here the same column density caps as described in
\S 3.2. Lines above $10^{16}$ cm$^{-2}$ for \ion{H}{i} and $10^{15}$
cm$^{-2}$ for metal lines are reset to $10^{16}$ cm$^{-2}$and
$10^{15}$ cm$^{-2}$, respectively.  While this particular statistic
has yet to be determined observationally as a function of impact
parameter, this could certainly be done, and would provide a robust
quantitative estimate of the total amount of absorption as one moves
away from galaxies.

The trends seen here in physical space are qualitatively similar
to those seen in redshift space in the previous section.  All the
ions show enhanced absorption near galaxies; even at 1~Mpc the
${dN}_{ion}/{dz}$ for targeted LOS is noticeably higher than that
for random LOS, shown as the diamond at 1~Mpc in each plot.  For
low and mid ionisation potential ions, the decline is very steep
with increasing radius, and beyond a few hundred kpc, ${dN}_{ion}/{dz}$
is almost independent of impact parameter.  For \ion{Mg}{ii}, the
dN/dz value has dropped by two orders of magnitude by 100~kpc. For
mid ions, the sharp drops in dN/dz happen at slightly further impact
parameters, roughly 200-300~kpc. For high ionisation potential ions,
the decline is not as steep with impact parameter, but there is
still a clear enhancement within roughy 300~kpc; about a factor of
two for \ion{O}{vi} and about a factor of four for \ion{Ne}{viii}.

There are also trends with halo mass, although they are not strong
as in Figure~\ref{ewvel}.  The general trend is that there is
somewhat more absorption at higher halo masses, at most impact
parameters.  To first order, this reflects the increased gas density
both in and around larger halos.  There are some interesting
exceptions; for instance, \ion{O}{vi} shows less absorption within
200~kpc for galaxies in massive halos.  This may reflect temperatures
deep within group-sized halos that exceed the collisional ionisation
temperature of \ion{O}{vi}~\citep{dav08_xray} combined with densities
that exceed the photo-ionisational densities of \ion{O}{vi}
\citep{opp09}.  When one reaches 1~Mpc impact parameters, the
${dN}_{ion}/{dz}$ values for \ion{H}{i} and the high ionisation
potential ions for the various halo masses all have essentially
converged.

\begin{figure*}
\subfigure{\setlength{\epsfxsize}{0.79\textwidth}\epsfbox{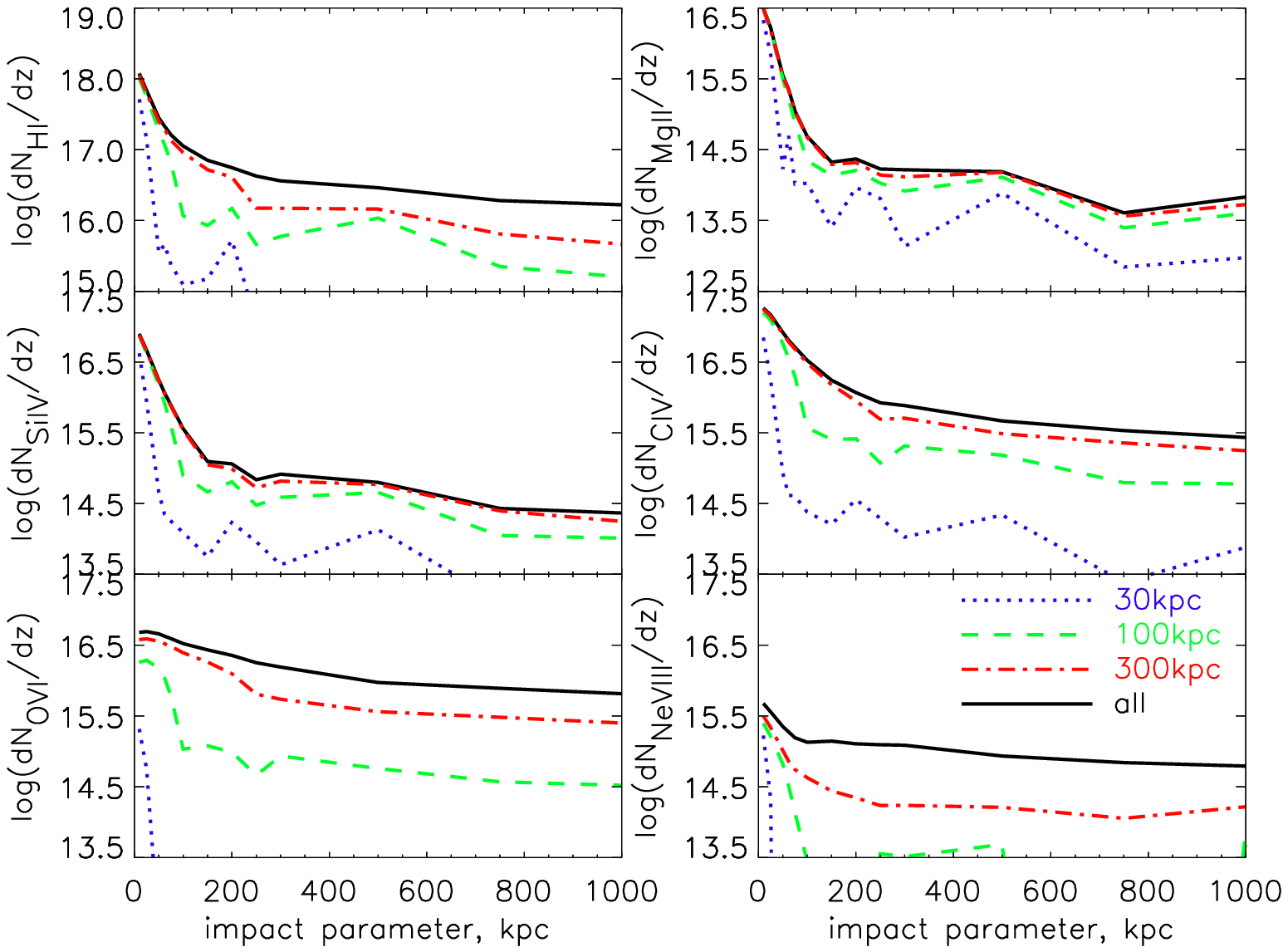}}
\caption{Total column density per unit redshift ($dN/dz$) for targeted LOS
around \mmid~halos, for all gas particles (black solid line, identical
to Figure \ref{NvsB}), and for gas within the specified radius of a
galaxy (coloured lines).  Black line is labeled "all", indicating that
we consider absorption from all SPH particles (i.e. the full simulation)
intersecting the targeted LOS. Coloured lines indicate $dN/dz$ where we
only include absorption from SPH particles that are within a sphere of
the indicated physical radius from {\it any} galaxy in our simulation (not
just the targeted one).  Dotted purple 
shows $dN/dz$ for gas within 30 kpc (physical) of a galaxy,
Dashed green line is for 100 kpc, and dash-dot red is for 300 kpc.}
\label{sel}
\end{figure*}

We have argued above somewhat indirectly that most metal absorption
arises from gas within roughly 300~kpc of galaxies (depending on
the ion), but in simulations we can test this hypothesis directly.
To do so, we tag gas particles in our simulations that lie within
a chosen radius $r$ from any resolved SKID-identified galaxy
(${M}_{*}\ge{\rm 10}^{9.1}~{M}_\odot$), and then regenerate our LOS
with contributions only from those tagged particles. We exclude
galaxies below this mass limit as they are not well resolved in our
simulations~\citep{fin06}.  Figure~\ref{sel} shows the resulting
summed ${dN}_{ion}/{dz}$ values as a function of impact parameter,
analogous to Figure~\ref{NvsB}, with the different lines showing
the contribution from particles within $r<30$~kpc (blue dotted),
$r<100$~kpc (green dashed), and $r<300$~kpc (red dot-dashed) from
galaxies.  We show here results for the \mmid\ halo mass bin, but
the results are not significantly different for other halo masses.
For comparison, the solid black line includes absorption from all
gas, reproduced from Figure \ref{NvsB}.

Explaining further, in this Figure we use the same galaxies as in
Figure \ref{NvsB}, with the same LOS at the same impact parameters.
But we no longer consider absorption from every SPH particle that
intersects the targeted LOS. We now consider absorption from the
SPH particles along the line of sight {\it only} if they are within
a given physical radius from {\it any} galaxy in our simulation
volume, {\it not just the targeted one.} The reader will note that,
for example, the red 300~kpc line extends past 300~kpc. This is due
to clustering. A targeted line of sight can find absorbers at, for
example, 750~kpc because those absorbers are from particles within
300~kpc of another galaxy, as opposed to the targeted one.

\begin{table}
\caption{For each ion, we give the percentage of absorption from particles within 300~kpc of any galaxy in our simulations, at each impact parameter listed. Values in this table are derived from the difference between the red dash-dotted and black solid lines in Figure \ref{sel} at impact parameters of 10, 100, and 300~kpc.}

\begin{tabular}{l c c c}
\hline
  &  10~kpc & 100~kpc & 300~kpc  \\ 
\ion{H}{i} & 94\% & 81\% & 41\%  \\ 
\ion{Mg}{ii} & 97\%& 97\% & 80\%  \\ 
\ion{Si}{iv} & 96\% & 97\% & 80\% \\ 
\ion{C}{IV} & 96\% & 92\% & 66\% \\
\ion{O}{vi} & 79\% & 74\% & 35\% \\ 
\ion{Ne}{viii} & 67\% & 32\% & 14\% \\
\hline

\end{tabular}
\label{table:percents}
\end{table}

For the low and mid metal ions, particles within 300~kpc are
responsible for the great majority of the absorption at all impact
parameters, as given in Table \ref{table:percents}. For low ionisation
species, particles within even smaller radii are responsible for
the majority of absorption. For \ion{Mg}{ii}, most of the absorption
owes to particles within only 100~kpc, and typically around half
of the \ion{Mg}{ii} absorption owes to gas within only 30~kpc of
galaxies.  For mid ions \ion{Si}{iv} and \ion{C}{iv}, essentially
all the absorption comes from within 300~kpc of galaxies, with the
majority of it from within 100~kpc, and only a small fraction within
30~kpc.  For \ion{O}{vi} and \ion{Ne}{viii}, how much absorption
comes from particles within 300~kpc depends upon impact parameter,
as shown in Table \ref{table:percents}. For these high ions, very
little of the absorption comes from particles within 100~kpc, and
in fact a substantial fraction can come from particles further than
300~kpc from galaxies.  The contributions to the 100~kpc curves at
${\it b} >$ 100~kpc, for example, must come from gas associated
with other galaxies, either satellite systems or galaxies projected
along the LOS with $| \Delta v |< 300 \kms$.

In summary, the extent of metal ions around galaxies -- that is,
the range containing the vast majority of metal absorption -- is
roughly 300 $\kms$ in velocity space (although in some cases much
less than that), and roughly 300~kpc in physical separation (except
for \ion{O}{vi} and \ion{Ne}{viii} at large impact parameter). Low
ionisation potential metal ions are more confined around galaxies
in both velocity and physical space. This is consistent with the
visual impression of the column density images in Figure~\ref{snap300}.
We reiterate that metal ions are more likely to be found within
300~kpc or 300 $\kms$ of {\it some} galaxy, not necessarily the
target galaxy. In particular, the excess of low ions at large radii
in Figures \ref{NvsB} and \ref{sel} arises from satellites and
neighboring galaxies.

\section{Column Density Distributions}
\label{sec:CDD}

\begin{figure*} 
\subfigure{\setlength{\epsfxsize}{0.79\textwidth}\epsfbox{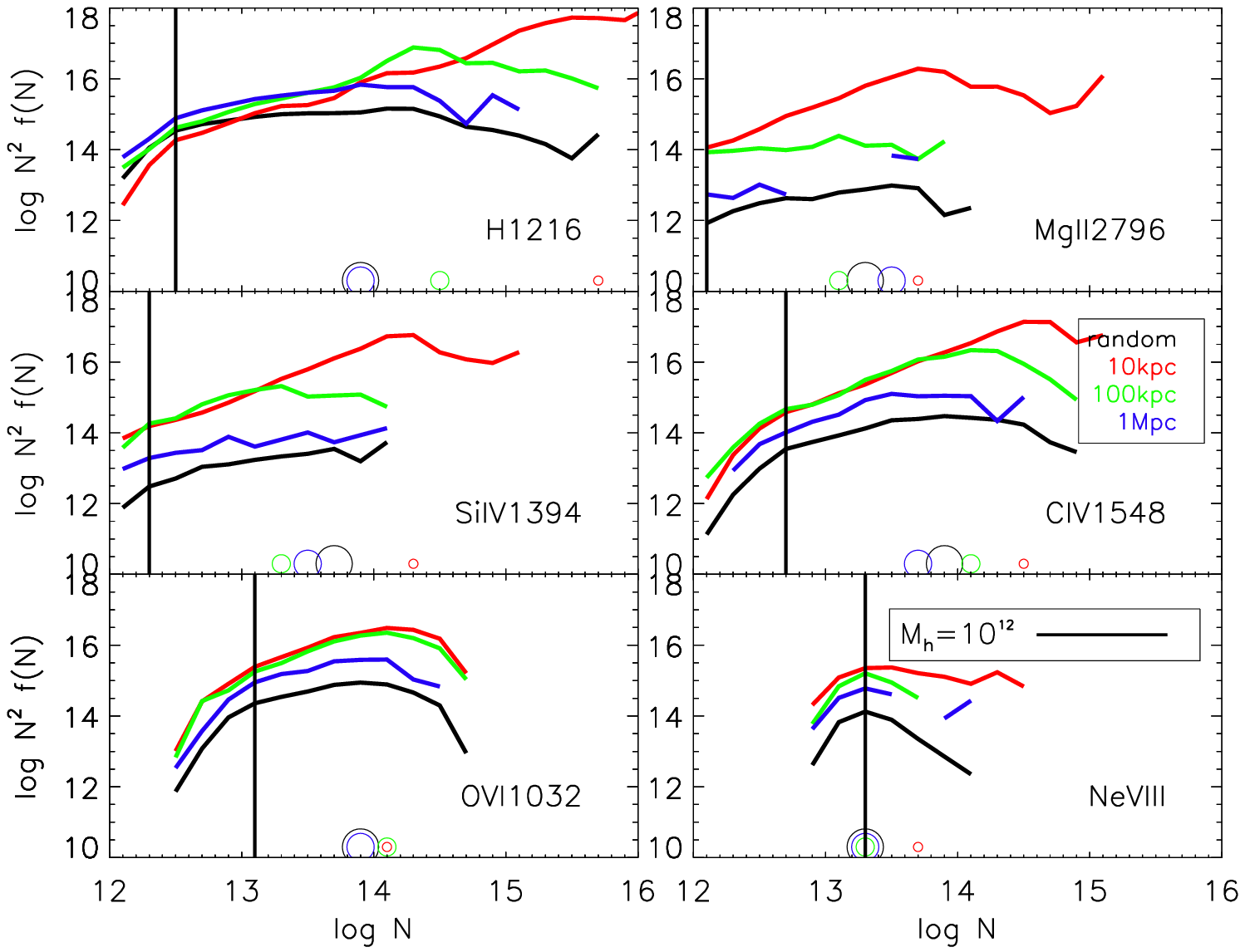}}
\subfigure{\setlength{\epsfxsize}{0.79\textwidth}\epsfbox{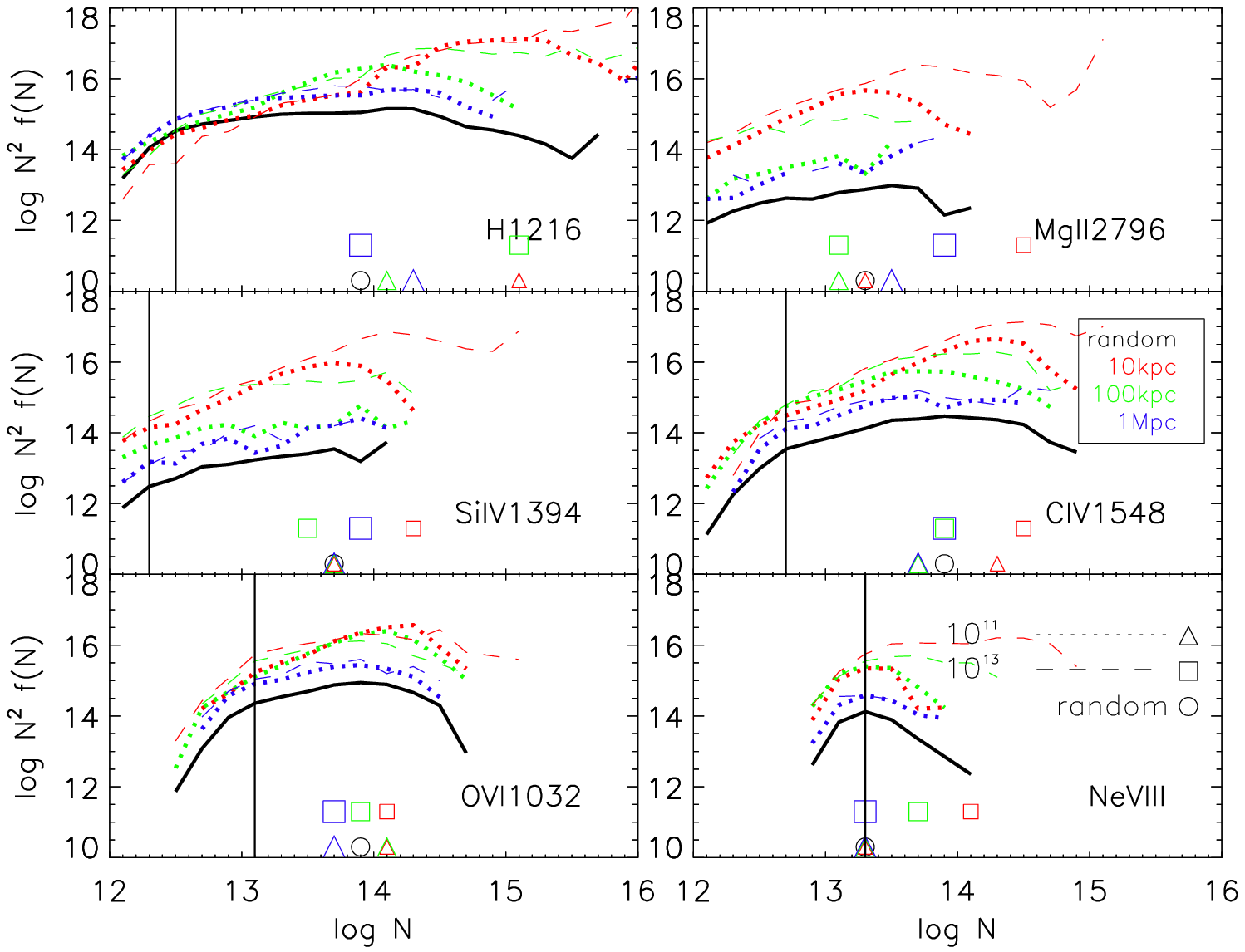}}
\caption{Top panel: Column density distributions (CDDs) for galaxies
with ${\rm M}_{halo}=10^{12} M_\odot$. ~${\it f}({\rm N})\equiv{\rm
d}^{2}n/{\rm d}{\it N}{\rm d}{\it z}$.  We multiply ${\it f}({\rm N})$
by ${\it N}^{2}$ to show from where most of the absorption per unit log
column density arises. The black line is the column density distribution
for random LOS,  red, green, and blue are for targeted LOS at 10~kpc,
100~kpc, and 1~Mpc, respectively. Circles (which increase in size
with increasing ${\it b}$) mark the peak of the distribution, i.e.,
where 50\% of the column density is below that value and 50\% above it.
Bottom: Same as the top panel for galaxies in halos of $10^{11} M_\odot$
(dotted line) and $10^{13} M_\odot$ (dashed line). Triangles and squares
mark the peak of the distribution for \mlo~and \mhi~respectively,
also increasing in size with increasing ${\it b}$, while circles mark
peak for random distribution, as in top panel. Vertical lines show the
completeness limit of the CDDs for S/N=30 per $6 \kms$ pixel.}
\label{cdd}
\end{figure*}

The column density distribution (CDD), i.e. the number of absorption
systems per unit column density per unit redshift, represents the
most basic counting statistic for characterising absorption line
systems.  Examining the CDDs as a function of impact parameter tells
us how absorption drops off with distance as a function of column
density, and thereby gives us a more detailed view of how absorption
varies around galaxies compared to the aggregate statistics presented
in the previous sections.

Figure~\ref{cdd} shows CDDs plotted as ${N}^{2} \times f(N)$ where
$f(N)\equiv d^{2}n/d N dz$ for all our ions, where $N$ is the column
density, $n$ is the number of lines, and $dz$ is the redshift-space
path length, corresponding to $\pm 300 \kms$.  We plot this at three
impact parameters of 10~kpc (red), 100~kpc (green), and 1~Mpc (blue),
and for comparison we also plot the CDD for random LOS~\citep[black;
similar to those in][]{opp12}.  As before, we combine ions with
separations $\Delta v < 100 \kms$ into systems before measuring the
CDD, and apply column density caps (see \S 3.2). In the top panel,
we add colored circles indicating the center of the distribution,
i.e., where 50\% of the total column density is below that value
and 50\% above it.  In the bottom panel we indicate these central
values with triangles and squares for \mlo~and \mhi, respectively.
The top panels show the results for galaxies in halos of $\approx
{10}^{12} M_\odot$.  The bottom panels show analogous CDDs for
galaxies in halos of $\approx {10}^{11} M_\odot$ (dotted lines) and
$\approx {10}^{13} M_\odot$ (dashed) to assess the dependence on
halo mass.

We multiply $f(N)$ by ${N}^{2}$ to obtain a quantity that reflects
the (relative) amount of absorption per unit redshift at each column
density in that ion: the value of $N^2 \times f(N)$ at $N=10^{14}\,{\rm
cm}^{-2}$, for example, represents the total column density per
unit redshift contributed by lines in a $\Delta\ln N=1$ interval
centered at $N=10^{14}\,{\rm cm}^{-2}$. Multiplying by $N^2$ also
enhances the visibility of trends by mitigating the typically steep
power-law dependence of $f(N)$.  In previous work \citep{opp12,dav10},
we had plotted ${N}\times f(N)$ just for visibility's sake.  Here
we add an additional power of $N$, because this means that the crest
of the ${N}^{2} \times f(N)$ curve represents the column density
contributing the most absorption per logarithmic interval in that
ion. These crests are typically shallow --- indicating absorption
that is spread over a fairly wide range of column densities --- but
they correspond well to the centers of the cumulative distributions
marked by the circles.  

We plot a vertical line that represents the 50\% completeness limit
for the CDD of absorbers identified in our artificial spectra with
S/N=30.  We determine this limit by comparing these CDDs to those
derived from artificial spectra with S/N=100 (not shown) and by
identifying the column density where the S/N=30 CDD begins to deviate
by more than 50\% from the CDD derived with S/N=100.  We determine
this limit using random LOS since this sample contains the largest
total number of lines owing to its large path length.  Since we
employ the same S/N in all spectra, this completeness limit should
also be applicable for the targeted LOS.

The first thing that one gleans from these figures is that for the
metals, even at impact parameters of 1~Mpc there is clearly more
absorption near galaxies than in the random LOS. This is true for
every ion and for the full range of halo masses that we explore in
this paper.  For low ions that arise predominantly near galaxies,
this reflects the large-scale auto correlation function of galaxies,
which extends to many megaparsecs.  For high ionisation potential
ions, the absorption arises in less dense (diffuse and WHIM) gas
that still correlates with galaxies living in large-scale structures.

We also see that the cosmic absorption in every metal ion peaks in
the range of $N\approx 10^{13}-10^{14.5} \cdunits$.  These peaks
are generally above the completeness limit for all the ions except
\ion{Ne}{viii}, where the lines are intrinsically weak. For the
lower ionisation potential ions (\ion{Mg}{ii}, \ion{Si}{iv},
\ion{C}{iv}), this peak moves to larger columns as one goes to
smaller impact parameters, while for the high ionisation potential
ions the peak remains mostly independent of the impact parameter.
In addition, for high ionisation potential ions (and for \ion{Mg}{ii})
one must probe to fairly low column densities to capture the bulk
of the cosmic absorption, e.g., $N_{\rm OVI}\sim 10^{14}\cdunits$
and $N_{\rm NeVIII}\la 10^{13.5}\cdunits$.  Hence, while many
\ion{O}{vi} and \ion{Ne}{viii} absorption systems are now being
detected in COS data~\citep{tum11,tri11}, our models predict that
higher S/N observations that probe to $N \sim 10^{13.5}\cdunits$
are needed to capture the bulk of cosmic absorption in these ions.

Now let us examine \ion{H}{i}, as this ion is relatively well
understood in terms of its connection with galaxies and large-scale
structure~\citep[e.g.,][]{dav99,dav10}.  \ion{H}{i} shows a clear
trend of having an increasing number of high column density absorption
systems as one approaches a galaxy, i.e. as one goes to small impact
parameters.  In contrast, as one goes to smaller impact parameters
the number of lower column density absorption systems does not
change as significantly.  In fact, below $N \sim 10^{14}\cdunits$
the trend actually reverses with the number of systems increasing
as one goes to larger impact parameters.  This can be easily
understood.  The strong absorbers mostly arise in the dense gas
around galaxies, while the weak absorbers arise more in the surrounding
large-scale structure, even when the impact parameter is small.
This is consistent with the strong correlation between \ion{H}{i}
column density and overdensity predicted in the
models~\citep[e.g.,][]{dav01,sch01,dav10,hui97}.

For the metal lines, the variation in the shape of the CDD with
impact parameter shows interesting trends as a function of ionisation
level. As one moves from impact parameters of 100 kpc to 10 kpc,
for lower column density absorbers there is little change in the
incidence of absorption.  However, as one moves to higher column
densities, the number of absorbers increases and the peak moves to
higher column densities.  The column density where the increase to
small impact parameter becomes noticeable increases with ionisation
level.  For instance, for \ion{Mg}{ii} the 10~kpc and 100~kpc curves
differ at $N_{\rm MgII}\ga 10^{12}\cdunits$, while for \ion{C}{iv}
they differ at $N_{\rm CIV}\ga 10^{14}\cdunits$.  For \ion{O}{vi}
and \ion{Ne}{viii}, there are only minimal differences in the CDDs
at impact parameters of 10 kpc and 100 kpc.  At impact parameters
of 1 Mpc the number of absorbers is smaller at all columns for all
the metal ions, with the peak also generally occurring at lower
column densities. The random LOS continue this trend.  In short,
the lower the ionisation potential, the more sensitive the ion's
CDD is to the proximity of a galaxy.

It is conventional to fit the CDD with a power law in column density
for a given ion: $f(N)\propto N^{-\beta}$.  For random lines of
sight, the observed power-law slopes for \ion{H}{i}, \ion{O}{vi},
and \ion{C}{iv} are generally between $1.5\la\beta\la
2.2$~\citep{dan06,coo10}.  \citet{dav10}, using similar simulations,
found a slope for the \ion{H}{i} CDD of $\beta=1.70$, which is
(unsurprisingly) similar to what we find here.  The slope for
\ion{O}{vi} depends on the range of column densities over which the
fit is done, since a pure power law is not a good descriptor.  Using
the full range above our completeness limit, we obtain $\beta\approx
2.3$, which is comparable to \citet{dan06} who found $\beta=2.2\pm
0.1$.  \ion{C}{iv} likewise is not a perfect power law, but we find
a slope of $\beta\approx 1.9$, which is steeper that observations
by \citet{coo10} that yield $\beta=1.5^{+0.17}_{-0.19}$, but probably
within uncertainties given that incompleteness in the data which
has poorer quality than our simulated spectra will generically lead
to shallower slopes.  Given the variations in spectral quality
between all these data sets, we consider our simulations to be
broadly in agreement with current measures of CDD slopes for these
ions.  Observations have yet to constrain the slope as a function
of impact parameter, but our models predict that the slope does not
vary dramatically, and mainly only the amplitude increases substantially
to small impact parameter.  This prediction will be testable with
upcoming observations that have sufficient statistics to examine
$f(N)$ as a function of impact parameter.

Looking at the bottom panels of Figure~\ref{cdd}, absorption around
galaxies in halos of larger and smaller masses continues these same
general trends.  At impact parameters of 10 kpc, as one goes from
halo with masses of $10^{11} M_\odot$ to $10^{13} M_\odot$ one sees
similar trends as when one went from impact parameters of 100 kpc
to 10 kpc in halos of fixed mass; there is agreement of the CDDs
at small column densities and more absorption at high column
densities, and the transition occurs at a column density that
increases as the ionisation potential of the ion increases.  Hence
in this regard, going to higher halo masses is equivalent to going
to smaller impact parameters in a halo of fixed mass.  The CDDs are
relatively insensitive to the mass of the galaxy halo for impact
parameters of 1 Mpc.

\hskip0.0001in \ion{Ne}{viii} exhibits an interesting trend at high
halo masses, arising from its strong collisional ionisation
contribution when a hot gaseous atmosphere is present.  We see in
Figure \ref{color_high} that \ion{Ne}{viii} has a low-density,
photo-ionised component probing the diffuse IGM, which gives it the
more extended characteristics of a high ionisation potential ion
like \ion{O}{vi} that is itself mostly photo-ionised.  However,
\ion{Ne}{viii} also traces $10^{5.5-6.0}$~K hot halo gas, which is
both denser and at lower impact parameters, where cooler ions like
\ion{Si}{iv} and \ion{C}{iv} are also found.  Therefore for large
halos and impact parameters inside of 100~kpc, there is a substantial
population of (collisionally ionised) \ion{Ne}{viii} absorbers.
Note that this \ion{Ne}{viii} could be coincident with lower
ionisation potential ions, as has been observed by \citet{tri11},
but arises in a different gas component.

In summary, the CDD of \ion{H}{i} and metal ions shows trends with
impact parameter that reflect correlations of absorption with both
nearby galaxies and large-scale structure.  All ions show increased
absorption closer to galaxies.  Low ionisation potential ions are
more influenced by the presence of nearby galaxies.  These CDDs
represent a prediction of hierarchical models that enrich the IGM
using outflows from star-forming galaxies and can in principle be
tested and constrained by observations~\citep[e.g.,][]{tum11}. The
full COS-Halos data set is being analyzed now (Werk et al., 2012),
and in future work we will undertake a detailed comparison to CDD
and other observed statistics, including an artificial spectra
sample that more closely mimics the COS-Halos spectra.

\section{Variations with outflow model}
\label{sec:model}

\begin{figure*} 
\subfigure{\setlength{\epsfxsize}{0.79\textwidth}\epsfbox{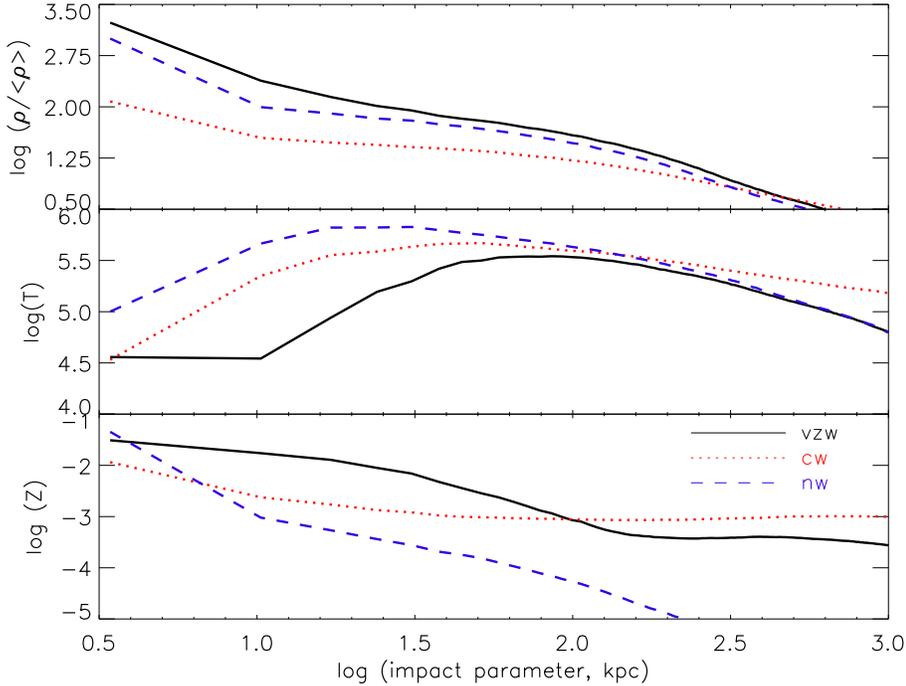}}
\caption{Median overdensity, temperature, and metallicity vs. impact parameter for
three different wind models: momentum-driven ``vzw", constant wind ``cw",
and no wind ``nw".  Sight lines are all around galaxies with halo masses
of \mmid. The top two rows are similar to the first two columns of Figure \ref{snap300}, but collapsing the image down to one dimension.}
\label{snap_model}
\end{figure*}

One expects the enrichment of the surrounding CGM gas to depend
sensitively on the properties of the enriching outflows.  So far
we have only considered our favoured outflow model with momentum-driven
wind scalings.  Here we consider how the absorption properties
around galaxies depend on our assumed outflow model, using two other
wind prescriptions: a simulation with no winds (nw) and a constant
wind (cw) model where we assume a constant mass loading factor of
$\eta=2$ and a constant wind speed from all galaxies of $v_w=680\kms$.
The latter is similar to that used in the Overwhelmingly Large
Simulations (OWLS) reference model of \cite {sch10}. The cw version
used in this paper is identical to that described in \cite{dav10},
except that it was re-run using \cite{wie09a} metal-line cooling
rates, for consistency with the vzw model used here.

To begin, Figure~\ref{snap300} shows a pictorial representation of
how the physical conditions and absorption vary with wind model.
In Figure \ref{snap_model}, for easier comparison, we collapse the
density and temperature information given in Figure~\ref{snap300}
down into one dimension by taking the azimuthal average of those
images.  In Figure~\ref{snap_model}, we show a larger region than
in Figure~\ref{snap300} to illustrate larger-scale trends, and
include metallicity to more fully understand the cooling processes.

In the top panel of Figure \ref{snap_model}, vzw curiously shows
more similarities to the no wind model (nw) than to the constant
wind model (cw).  This illustrates that the high wind speeds from
all galaxies in the cw model cause significantly more spatial
dispersal of mass (along with metals, as we show below) on $\sim$Mpc
scales around galaxies.  The vzw model has lower velocity winds
that do not have such a dramatic impact on large scales, but do
have a strong impact on smaller (CGM) scales. The vzw model pushes
more mass via winds into the CGM relative to the no wind model, but
this mass is close enough that it recycles back onto the galaxy in
much less than a Hubble time \citep{opp10}. We note that the galaxy
population is broadly more similar in the two wind models, as they
both suppress global star formation substantially relative to the
no-wind case~\citep[for a fuller discussion of these properties,
see][]{opp10,dav11a,dav11b}, but Figure~\ref{snap_model} shows that
they are quite different in where they deposit the ejected material.

The middle and lower panels of Figure \ref{snap_model} illustrate
how winds affect the CGM.  Here vzw and cw actually trend in the
opposite direction relative to the no-wind case! In the vzw case,
the area around the galaxy is slightly colder than with no winds.
This is because the lower wind speeds deposit more metals around
galaxies (lower panel), and this results in an increased amount of
metal cooling~\citep{opp12} that more than offsets the shock heating
from the winds.  In contrast, the cw model expels gas at high
velocities, around the escape velocity for \mmid\ halos, even from
small galaxies. This means the enrichment is more widespread --
past 100~kpc, cw shows greater metallicity than the vzw or nw model.
Moreover, the wind energy is deposited into less dense gas where
it does not have a chance to radiate away its energy.

The \ion{H}{i} maps shown in Figure \ref{snap300} illustrate that
winds do not make a large difference to \ion{H}{i} absorption, at
least on the large scales depicted here, similar to results at high
redshift \citep[e.g.,][]{kol03,kol06}.  However, there is an increase
in \ion{H}{i} on small scales, roughly 100~kpc. The metal absorption,
in contrast, shows more dramatic differences between the wind models.
In the no-wind case, all the metals are essentially confined to be
in and around galaxies, showing that even the distribution of metals
within the CGM requires winds. These differences should be manifest
in the statistics of absorbers around galaxies, providing an
opportunity to constrain wind models.

\begin{figure}
\subfigure{\setlength{\epsfxsize}{0.5\textwidth}\epsfbox{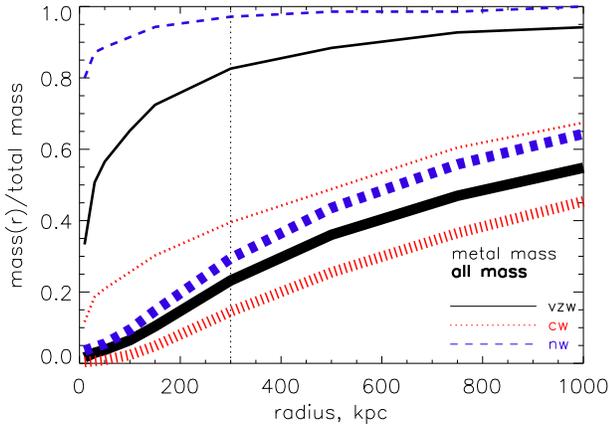}}
\caption{Thin lines: Metal mass fraction of all gas particles within spheres of radius $r$ around all galaxies ${M}_{*}\ge{\rm 10}^{9.1}~{M}_\odot$
in our simulation at z=0.25, relative to the total metal mass of gas particles in the whole simulation
volume, for momentum-driven winds (solid line), no winds (dashed line), and 
constant winds (dotted line).  The vertical dotted line delineates 300~kpc.
Thick lines: Total mass fraction for all species, not just metals. }
\label{massfrac}
\end{figure}

To further quantify the extent of the metal distribution, Figure
\ref{massfrac} shows the fraction of all cosmic metal mass that
lies within a given radius from galaxies with ${M}_{*}\ge{\rm
10}^{9.1}~{M}_\odot$ in the simulation volume. We also plot the
fraction of all cosmic mass for all species, in addition to just
metals. The procedure for this was discussed in \S\ref{sec:b}.  Here
the impact of winds on distributing cosmic metals is shown clearly:
the no wind model keeps essentially all metals confined very close
to galaxies, while the constant wind model disperses them over large
scales, with the momentum-driven wind model intermediate between
the two.  This figure also shows that the constant wind model also
pushes the total mass further out, while the momentum-driven wind
and no wind models are more similar. For our favoured momentum-driven
wind scaling model, 83\% of the metal mass is in gas within 300~kpc
from galaxies.  For the cw model, only 40\% is within this radius.
Hence the basic extent of metals around galaxies can already provide
a discriminant between wind models.

\begin{figure*} 
\subfigure{\setlength{\epsfxsize}{0.79\textwidth}\epsfbox{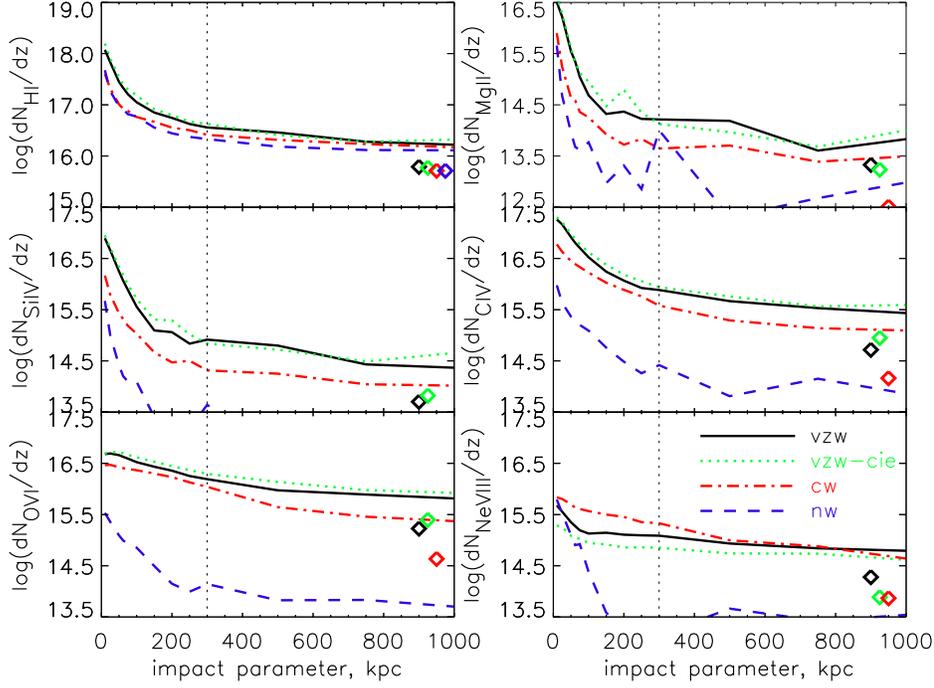}}
 \caption{The total column density per unit redshift for the
vzw, vzw-cie, cw, and nw models, all for galaxies in $10^{12}
M_\odot$ halos.
The diamonds indicate the ${dN}_{ion}/dz$ value for random lines of sight for
each wind model. The symbols are shifted slightly to the left of 1 Mpc, and
separated for easier viewing. In some cases the symbols are lower than the
scale of the plot so do not appear.}
\label{NvsB_wind}
\end{figure*}

We now aim to quantify these differences in metal distribution using
absorption line statistics.  Figure~\ref{NvsB_wind} is similar to
Figure~\ref{NvsB}, except that it compares the three wind models,
using the same column density caps as explained in \S 3.2.  Here,
we also include a comparison to our old vzw simulation that used
collisionally ionised equilibrium (CIE) metal line cooling, labeled
as vzw-cie, as opposed to our current vzw model that uses the
\citet{wie09a} (PIE) cooling.  For the most part, the differences
between the vzw-cie model and our current vzw model are small
compared to the differences between wind models, and hence this
aspect of our modeling does not introduce a large uncertainty into
the results. The differences between vzw and vzw-cie are small
because the winds have moderate velocities, preventing large amounts
of gas to reach lower overdensities. Instead, they remain in
moderate-density gas where cooling times are short regardless of
photo-ionisation suppression. Note that when we generate spectra,
we compute ionic abundances including photoionisation in both cases;
it is only during the evolution of the simulation that vzw-cie is
different.

In this figure, as in Figure \ref{NvsB}, we also show ${dN}_{ion}/dz$
for random LOS for each wind model as the diamonds near the right
edge. For all models and ions, ${dN}_{ion}/dz$ values at 1~Mpc are
still higher than ${dN}_{ion}/dz$ for random LOS.  Note that for
all the metal ions, there is so little absorption in the no wind
model that the random LOS ${dN}_{ion}/dz$ (purple diamonds) falls
off the bottom of this plot.  Similarly, the cw model shows so
little \ion{Si}{iv} and \ion{Mg}{ii} absorption in random LOS that
the red diamonds fall off the bottom of the plot.

For all the metal ions, the no wind model gives significantly less
absorption than any of the wind models, even down to the smallest
impact parameters probed here.  However, there is not a large
difference between models with winds and the model without winds
for \ion{H}{i}.  Without outflows, metals basically exist outside
the ISM only owing to tidal or ram pressure stripping processes
that remove material from the ISM of the central or satellite
galaxies.  It is evident from this plot (as with the images in
Figure~\ref{snap_model}) that such stripping processes provide only
a small contribution to the CGM metal absorption in these halos,
though it can be more substantial in large halos
~\cite[e.g.,][]{dav08_xray,Zu11}.  Hence, we predict metals seen
at any impact parameter beyond that of the ISM of typical galaxies
arise almost exclusively owing to outflows.

\begin{figure*}
\subfigure{\setlength{\epsfxsize}{0.79\textwidth}\epsfbox{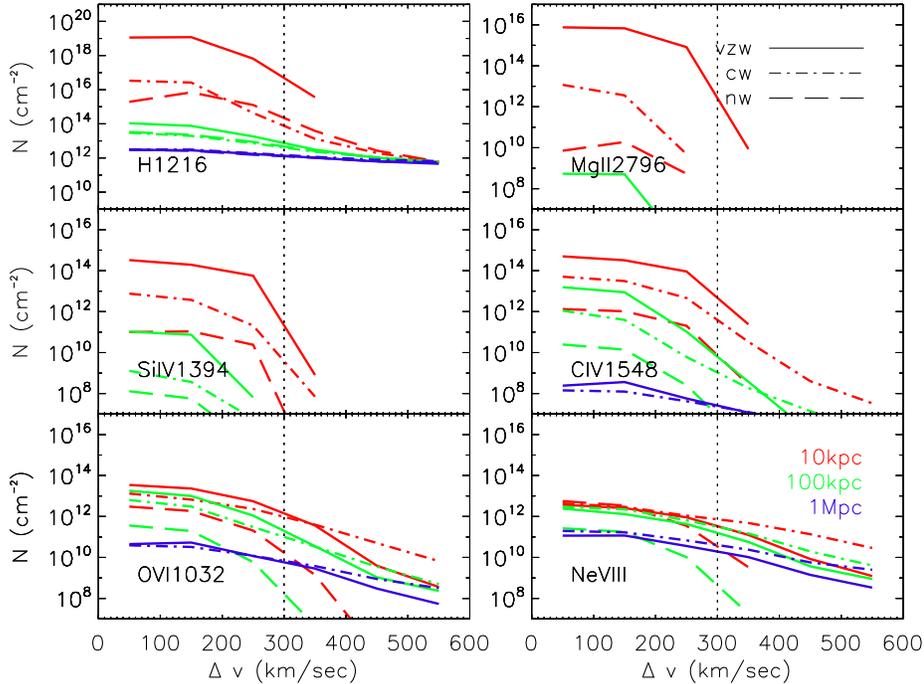}}
 \caption{The median column density (N) vs. velocity separation
from the central galaxy ($\Delta v$) for galaxies in 
${\rm 10}^{12}M_\odot$ halos for the vzw, cw, and nw models.
We show results at three impact parameters:
10 kpc (red), 100 kpc (green), and 1 Mpc (blue). The vertical dotted marks
$300 \kms$.}
\label{ewvelM}
\end{figure*}

The differences between the constant wind and momentum-driven wind
cases are more subtle.  The global trends with ionisation state are
as before: lower ionisation potential species are more highly peaked
at small impact parameters in both wind models.  The main significant
difference is an overall offset.  In general, the cw model has lower
absorption for low ionisation potential species, and higher absorption
for high ionisation potential species relative to the vzw (or
vzw-cie) model.  The lower absorption arises in part because the
cw model produces less metals overall owing to its reduced cosmic
star formation~\citep{opp12}.  Going to higher ionisation species,
absorption in the cw model becomes only slightly less than absorption
in the vzw model. For \ion{Ne}{viii}, cw shows more absorption than
vzw.  This reflects the impression from Figure~\ref{snap_model}
that the cw model expels metals to greater distances and heats this
diffuse gas more~\citep{opp12}.  Hence, the global absorption
strength as a function of ionisation state provides another potential
discriminant between outflow models.

In Figure \ref{ewvelM}, we plot the median column density versus
velocity, analogous to Figure \ref{ewvel} except now we vary the
wind model as opposed to the halo mass.  Not surprisingly, for
\ion{H}{i} all the wind models give similar results; however vzw
does show an increase over nw and cw at 100~kpc, because these winds
push more cool gas to CGM distances. For the metal lines, once again
the no-wind case shows virtually no absorption except perhaps very
(dynamically) close to galaxies ($\pm 100 \kms$).  Hence, tidal
effects and other stripping processes are ineffective in distributing
metals in velocity space, just as in physical space.

In this figure, we see similar trends with wind model as we did in
Figure \ref{NvsB_wind}.  For low ionisation potential metal species,
the vzw model produces higher column densities, while the opposite
is true for high ionisation potential species.  Previous work
\citep{opp12} has shown that the cw and vzw models enrich the dense
gas in quite different ways. The cw model, with its high velocities
emanating even from small galaxies, deposits fewer metals into the
high density regions very close to galaxies and more metals into
the diffuse IGM.  Accordingly, metal ions that show more absorption
in the cw model than in the vzw model have more absorption from
more diffuse IGM gas.

In summary, the rate of the decrease in absorption of metal lines
as one moves away from the central galaxy provides a potentially
strong discriminant between outflow models that enrich the diffuse
IGM to fairly similar levels.  Winds are required to enrich the CGM
of normal galaxies to any significant level.  Our currently favoured
momentum-driven wind scaling model predicts more low ionisation
potential absorption close to galaxies than the constant wind model
and less high ionisation potential absorption farther from galaxies.
Quantitative comparisons with present and upcoming COS data should
yield more stringent constraints on outflow propagation.

\section{Conclusions}
\label{sec:conclusions}

We have examined the absorption line properties of \ion{H}{i} and five
key metal ions in the vicinity of galaxies at $z=0.25$ in cosmological
hydrodynamic simulations that include galactic outflows, as a function
of velocity separation, impact parameter between the galaxy and the
line of sight, and halo mass.  Our chosen metal ions span a range of
ionisation potentials, from low (\ion{Mg}{ii}) to mid (\ion{Si}{iv} \&
\ion{C}{iv}) to high (\ion{O}{vi} \& \ion{Ne}{viii}).  This is the first
absorption line study of metal absorption around galaxies in cosmological
hydrodynamic simulations with a self-consistently generated outflow
model that matches a wide range of galaxy and IGM observables.  Our work
is motivated by the upcoming wealth of data from {\it Hubble's} Cosmic
Origins Spectrograph, and hence our artificial spectra are generated
with resolutions and noise characteristics comparable to the best data
that will be obtained with COS.

Our primary conclusions are:

\begin{enumerate}

\item Absorption in all ions is enhanced closer to galaxies.  This
is true in terms of both impact parameter and velocity separation.
A velocity separation of $\pm300\kms$ around galaxies encompasses
most of the cosmic metal line absorption (although in some cases
it can be less), and absorption is also significantly stronger
within 300~kpc around a galaxy, particularly for lower ionisation
potential lines.  These ranges are somewhat larger around bigger
galaxies.  In our favoured momentum-driven wind simulation, $\approx$
80\% of all cosmic metals lie within 300~kpc of a galaxy.

\item The dependence of metal absorption strength on distance from
a galaxy in either physical or redshift space depends monotonically
on the ionisation potential of the absorbing ion.  Ions with a low
ionisation potential (\ion{Mg}{ii}, \ion{Si}{iv}) arise in higher
density gas that tends to drop off more quickly with impact parameter
and velocity separation than ions with higher ionisation potentials
(\ion{C}{iv}, \ion{O}{vi}, \ion{Ne}{viii}).  High ionisation potential
absorbers are more associated with gas at halo-like overdensities,
and the overall large-scale structure that contains galaxies, while
low ionisation potential lines arise more in the dense gas close
to individual galaxies.  Regardless of ionisation level, even out
to 1~Mpc, targeted LOS show an excess of absorption over random
LOS, reflective of large-scale matter clustering. 

\item The majority of cosmic absorption in the ions considered here
occurs in photo-ionised gas at $T<10^5$~K.  The exception to this
is for the high ionisation lines \ion{O}{vi} and \ion{Ne}{viii} at
impact parameters $\la 100$~kpc arising in massive halos containing
substantial hot gas; these lines are predominantly collisionally
ionised.

\item The dependence of the column density distributions on impact
parameter also shows trends with ionisation potential.  Species
with lower ionisation potentials are more affected by the proximity
of a galaxy.  Excepting \ion{Ne}{viii}, our spectra (comparable to
COS data quality)  directly trace the majority of cosmic absorption;
i.e. there is not a large population of smaller absorbers that is
inaccessible to COS that would dominate the total cosmic absorption.

\item Without winds, even gas within $\approx$ 100~kpc of galaxies
remains mostly unenriched; hence outflows are required to enrich
the CGM as well as the IGM.  The differences in metal absorption
between our favoured momentum-driven wind scaling model and a
constant wind model are evident both for lower ionisation potential
species close to galaxies and for high ionisation potential species
farther from galaxies. The models are clearly distinguishable from
each other using the combined CGM statistics of \ion{O}{vi} and
\ion{H}{i}.

\end{enumerate}

These results provide a starting point for understanding how
absorption lines trace the metal enrichment around galaxies.  We
have highlighted some basic trends, and shown that quantifying this
distribution as a function of impact parameter and velocity separation
can provide interesting constraints on key physical processes such
as galactic outflows, as well as the density and temperature state
of the metal-enriched IGM.

While we have considered only five metal ions in this paper, the
basically monotonic trends with ionisation potential suggest that
the behaviour of any other ion can be predicted just based on its
ionisation potential.  Our simulations predict that higher ionisation
potential ions should fall off more slowly at larger impact parameters
and velocity separations, because they are tracing lower density,
mostly photo-ionised gas.  This suggests a physical structure of
CGM gas in which \ion{O}{vi} is more extended than \ion{C}{iv},
which is more extended than \ion{Si}{iv}, and so on.  \ion{Ne}{viii}
and \ion{O}{vi}, if probed at sufficiently low column densities,
trace the most remote and diffuse metals of any UV resonance line,
but their strongest lines arise in hot collisionally ionised gas
near large galaxies.

This work is the first step in a series of works to confront
successful models for galaxy-IGM coevolution with absorption line
observations around galaxies in the low-z universe.  We examine
redshift $z=0.25$, but the basic trends are applicable to all
redshifts probed by COS ($z\la 1$), as recent work by \cite{dav10}
and \cite{opp12} show little evolution in the IGM from $z=1\rightarrow
0$.  We caution, though, that detailed comparisons to observations
are premature for the simulated spectra presented here, since
quantitative trends can be sensitive to the details of spectral
resolution, noise level, etc. Furthermore, our massive galaxies
($M_{\rm halo}= 10^{13} \msolar$) are all star-forming in these
simulations, in clear conflict with observations of mostly passive
galaxies in this halo mass range, and hence the predictions for
such halos may be influenced by physical effects not included in
our current models.  Nonetheless, we believe that the basic intuition
of understanding metal absorption surrounding galaxies in terms of
the ionisation potential of the tracer ion is robust, and provides
a clear intuition for interpreting current and future observations.
As observations progress, particularly with COS, simulations like
these will provide a critical testbed for galaxy formation models,
and will help elucidate the physical processes that drive the
enrichment of the intergalactic and circumgalactic medium.

\section{Acknowledgements}

We thank Jason Tumlinson, Todd Tripp, Molly Peeples, Joop Schaye,
Ben Weiner, Greg Walth, and Greg Stinson for useful discussions.
Partial support for this work came from NASA ATP grant NNX10AJ95G,
HST grants HST-GO-11598 and HST-GO-12248, NASA ADP grant NNX08AJ44G,
and NSF grants AST-0847667, AST-0907998, and AST-133514.  The
simulations used here were run on the University of Arizona's SGI
cluster, ICE, and on computing facilities owned by the Carnegie
Observatories. Computing resources used for this work were made
possible by a grant from the the Ahmanson foundation, and through
grant DMS-0619881 from the National Science Foundation.

\clearpage

\end{document}